\documentclass[12pt]{iopart}
\usepackage{rotating,graphics,graphicx,subfigure,amssymb}

\begin{document}

\title[Electric double layers with modulated surface charge density]{Electric
double layers with modulated surface charge density: Exact 2D results}

\author{Ladislav \v{S}amaj}

\address{Institute of Physics, Slovak Academy of Sciences, 
D\'ubravsk\'a cesta 9, 84511 Bratislava, Slovakia}
\ead{Ladislav.Samaj@savba.sk}
\vspace{10pt}
\begin{indented}
\item[]
\end{indented}

\begin{abstract}
Electric double layers (EDL) with counterions only, say electrons
with the elementary charge $-e$, in thermal equilibrium at the inverse
temperature $\beta$ are considered. 
In particular, we study the effect of the surface charge
modulation on the particle number density profile and the pressure.
The mobile particles are constrained to the surface of a 2D cylinder and
immersed in vacuum (no dielectric image charges).
An EDL corresponds to the end-circle of the cylinder which carries
a (fixed) position-dependent line charge density.
The geometries of one single EDL and two EDLs at distance $d$ are
considered; the particle density profile is studied for both geometries,
the effective interaction of two EDLs is given by the particle pressure
on either of the line walls.
For any coupling constant $\Gamma\equiv \beta e^2 = 2\times {\rm integer}$,
there exists a mapping of the 2D one-component Coulomb system onto
the 1D interacting anticommuting-field theory defined on a chain of sites.
Using specific transformation symmetries of anticommuting variables,
the contact value theorem is generalized to the EDL with modulated line charge
density.
For the free-fermion coupling $\Gamma=2$ it is shown that, under
certain conditions, the matrix of interaction strengths between anticommuting
variables diagonalizes itself which permits one to obtain exact formulas for
the particle density profile as well as the pressure.
The obtained results confirm the previous indications of weak-coupling
and Monte Carlo observations that the surface charge inhomogeneity implies
an enhancement of the counterion density at the contact with the charged line
and a diminution of the pressure between two parallel lines in comparison
with the uniformly charged ones (with the same mean charge densities).
\end{abstract}


\vspace{2pc}

\submitto{\JPA}

\maketitle

\renewcommand{\theequation}{1.\arabic{equation}}
\setcounter{equation}{0}

\section{Introduction} \label{Section1}
Equilibrium statistical mechanics of classical particle systems interacting
pairwisely via the Coulomb potential is of particular interest in soft matter
and condensed matter physics.
In Gauss units, the three-dimensional (3D) Coulomb potential
in vacuum (dielectric constant $\varepsilon=1$) has the form
$\phi({\bf r}) = 1/r$ where $r$ is the modulus of ${\bf r}$. 
The definition of the Coulomb potential can be extended to any 
Euclidean space of dimension $\nu=1,2,\ldots$
as the solution of the Poisson equation
\begin{equation} \label{Poisson}
\Delta \phi({\bf r}) = - s_{\nu} \delta({\bf r}) ,
\end{equation}
where $s_{\nu}=2\pi^{\nu/2}/\Gamma(\nu/2)$ is the surface area of
the $\nu$-dimensional unit sphere.
The Fourier component of $\phi({\bf r})$ exhibits the
singular behaviour of type $1/k^2$ which preserves many generic properties
of 3D Coulomb systems like screening and the corresponding sum rules
\cite{Martin88}.
In an infinite two-dimensional (2D) space, the solution of (\ref{Poisson}),
subject to the boundary condition $\nabla \phi({\bf r}) \to 0$
as $r\to\infty$, reads as $-\ln(r/L)$ where $L$ is a free length scale.
The system of 2D pointlike charges can be represented in the real 3D space
as parallel infinite charged lines perpendicular to the given surface,
the model which mimics 3D polyelectrolytes.
In one dimension (1D), the Coulomb potential is given by $-r$.

In this paper, we study one-component Coulomb models of mobile pointlike
particles of the same (say elementary) charge $-e$.
In contrast to the standard jellium systems, the neutralizing background
charge density is spread over boundary surfaces of the domain
the charged particles are confined to. 
This kind of models occurs in biological experiments with colloids
immersed in polar solvents like the water.
The colloid surface acquires a surface charge density by releasing
micro-ions into the polar solvent \cite{Andelman06,Levin02}.
The fixed surface charge density is opposite to the charge of mobile
particles which are therefore referred to as ``counterions''.
In theoretical studies, the curved surface of large colloids is usually
substituted by a flat surface and the modulated charge density on the
colloid's surface by the uniform one.

In the one-wall geometry, the charged surface and the surrounding counterions
in thermal equilibrium form a neutral entity known as the electric double
layer (EDL) \cite{Gulbrand84,Attard88,Attard96,Messina09}.
The particle number density at the flat surface is related to the wall's
uniform surface charge density via a simple contact-value theorem 
\cite{Henderson78,Henderson79,Blum81,Carnie81}.
As concerns the geometry of two parallel uniformly charged walls with
counterions in between, their effective interaction mediated by counterions
is of particular interest \cite{Hansen00}.
To obtain the effective interaction of the walls, one calculates
the pressure from counterion densities at the contact with either of the walls.
A counter-intuitive attraction of like-charged colloids was observed
at small enough temperatures, experimentally
\cite{Khan85,Kjellander88,Bloomfield91,Kekicheff93,Dubois98}
as well as by computer simulations
\cite{Gulbrand84,Kjellander84,Gronbech97}; for more recent progress
in the field, see reviews \cite{Boroudjerdi05,Naji13}.

The weak-coupling (WC) limit of models with uniformly charged surfaces
and counterions only is described by the mean-field Poisson-Boltzmann (PB)
theory \cite{Andelman06} and its systematic improvement via a
loop expansion \cite{Attard88,Netz00,Podgornik90}.
The strong-coupling (SC) limit is more controversial.
Based on a virial fugacity expansion, the leading SC term of
the particle number density corresponds to a single-particle theory in
the electric potential induced by charged wall(s)
\cite{Moreira00,Moreira01,Netz01}
which was confirmed by Monte Carlo (MC) simulations
\cite{Moreira00,Moreira01,Moreira02,Kanduc07}.
Next correction orders of the virial SC approach fail to reproduce
correctly MC data.
Another type of SC theories is based on the creation of classical
Wigner crystals on the wall surfaces at zero temperature
\cite{Shklovskii99,Levin99,Grosberg02}.
A harmonic expansion of the interaction energy in particle deviations 
from their ground-state Wigner positions \cite{Samaj11a,Samaj11b}
reproduces the leading single-particle theory and implies a first correction
term to the counterion density which is also in excellent agreement with
MC data, from strong up to intermediate Coulombic couplings.
The accuracy of the analytic results was improved by adapting the idea of
a correlation hole \cite{Nordholm84,Forsman04} into the Wigner-structure
description in \cite{Samaj16,Palaia18}.

The subject of interest in this paper is the effect of the space modulation
of the fixed surface charge density on plates on the density profile of
counterions and the effective interaction between two plates.
The discreteness of the surface charge is omnipresent in real systems
and often plays a dominant role in bio-interfacial phenomena
\cite{Israelachvili92,Leckband93}.
Numerous experimental studies are reviewed in \cite{Walz98}.
Early theoretical approaches were based on liquid-state theory
\cite{Chan80,Kjellander88b}.
Analytic perturbation techniques were combined with MC simulations
to describe the effect of surface charge inhomogeneities in
the WC \cite{Lukatsky02a,Henle04} and SC \cite{Fleck05} regimes.
The obtained results indicate an enhancement of the counterion density
close the charge-modulated surface in comparison with the uniformly charged
one (with the same mean charge density).   
As concerns the geometry of two parallel inhomogeneously charged surfaces,
according to the PB theory the increase of the counterion density close
to surfaces means a decrease of counterions at the midplane and,
consequently, a diminution of the pressure \cite{Lukatsky02b,Khan05}.
Exact PB solutions of one planar or cylindrical EDL with the surface charge
modulation along one direction only were found recently by exploring in
an inverse way the general solution of the 2D Liouville equation \cite{Samaj19}.

As concerns the thermal equilibrium of 2D plasmas, the relevant parameter
is the coupling constant $\Gamma=\beta e^2/\varepsilon$
with $\beta=1/(k_{\rm B}T)$ being the inverse temperature and $\varepsilon$
(set to unity, for simplicity) the dielectric constant of the medium
the particles are immersed in.
2D Coulomb systems are of special interest because besides the PB limit
$\Gamma\to 0$ their thermodynamics and many-body densities are exactly
solvable also at the finite (free-fermion) coupling $\Gamma=2$,
in the bulk \cite{Jancovici81,Alastuey81} and in the semi-infinite as well as
fully finite geometries \cite{Jancovici92,Forrester98}.
Another advantage of the 2D one-component systems is an explicit
representation of their partition function and particle densities
for the sequence of the coupling constants $\Gamma = 2\gamma$ where
$\gamma$ is a positive integer (in the fluid regime),
expressing the integer powers of Vandermonde determinant by using
either Jack polynomials \cite{Tellez99,Tellez12} or anticommuting-field
theory defined on a one-dimensional (1D) chain of sites
\cite{Samaj95,Samaj04a,Samaj04b,Samaj15}.
In the case of uniform surface charge densities with counterions
only at $\Gamma=2$, the density profile of counterions for the one-wall
geometry was derived in \cite{Jancovici84,Samaj13} and the effective
interaction between two parallel asymmetrically charged plates at distance
$d$ in \cite{Samaj20}.

The aim of this work is to extend, within the anticommuting-field
representation, exact results for 2D one-component plasmas with uniformly
charged plates to plates with periodically modulated surface charge densities.
Using a cylinder geometry of the confining domain and applying certain
linear transformations of anticommuting variables leading to specific sum
rules, the contact value theorem for uniformly charged lines is generalized
to the lines with periodically modulated surface charge densities
at arbitrary coupling constant $\Gamma$. 
Under certain conditions, the periodic modulation of the surface charge
density does not destroy the exact solvability of the 2D
counterions systems at $\Gamma=2$.
Our exact results confirm previous WC and MC observations about the effect
of surface charge inhomogeneities on the enhancement of the density of
counterions at the wall contact and the diminution of the pressure between
two plates.

The paper is organized as follows.
In section \ref{Section2}, we present the general formalism of the canonical
ensemble for the system of identical charges confined to the surface of
a cylinder, for both cases of a single EDL (section \ref{Section2.1}) and
two parallel EDLs (section \ref{Section2.2}) with the modulated line
charge densities.
Section \ref{Section3} concerns general results obtained from the mapping
of the 2D one-component Coulomb system onto the anticommuting-field
theory on a 1D chain of sites.
The mapping for the coupling $\Gamma=2\gamma$ ($\gamma$ a positive integer)
is reviewed in section \ref{Section3.1}, section \ref{Section3.2} deals
with the exactly solvable free-fermion case $\Gamma=2$ and section
\ref{Section3.3} is devoted to the derivation of sum rules
for correlators of anticommuting variables.
Exact results for the geometry of one EDL are reported in section
\ref{Section4}.
The previously derived sum rules are used to generalize the contact value
theorem for uniformly charged lines to inhomogeneously charged lines
in section \ref{Section4.1}.
Special conditions under which the model is exactly solvable at the
free-fermion coupling $\Gamma=2$ are shown in section \ref{Section4.2}.
Exact results for the geometry of two parallel EDLs are reported in section
\ref{Section5}.
A counterpart of the contact value theorem is derived in section
\ref{Section5.1}, a general formula for the pressure in terms of the
counterion density profile valid for any coupling constant $\Gamma$
is given in section \ref{Section5.2}.
As concerns the free-fermion coupling constant $\Gamma=2$,
the counterion density profile is derived in section \ref{Section5.3}
and the pressure in section \ref{Section5.4}.
A short recapitulation is outlined in the concluding section \ref{Section6}.

\renewcommand{\theequation}{2.\arabic{equation}}
\setcounter{equation}{0}

\section{Cylinder geometry} \label{Section2}
Let $N$ mobile pointlike particles with the elementary charge $-e$ be confined
to the surface of a cylinder of circumference $W$.
The surface of the cylinder is equivalent to a 2D rectangle domain $\Lambda$
of points ${\bf r}=(x,y)$ with coordinates $y\in [0,W]$ and the periodic
boundary conditions at $y=0,W$.
The dielectric constant of the medium the particles are immersed in is equal
to that of vacuum, $\varepsilon=1$, and there are no dielectric image charges.

The Coulomb potential $\phi$ at a spatial position ${\bf r}\in\Lambda$, induced
by a unit charge at the origin ${\bf 0}$, is defined as the solution of 
the 2D Poisson equation (\ref{Poisson}) under the periodicity requirement
along the $y$-axis with period $W$.
Writing the potential as a discrete Fourier series in $y$, it reads as
\cite{Choquard81}  
\begin{equation} \label{periodicCoulomb}
\phi({\bf r}) = \frac{1}{W} \sum_{k_y} \int_{-\infty}^{\infty} 
{\rm d}k_x \frac{1}{k_x^2+k_y^2} {\rm e}^{{\rm i}(k_x x + k_y y)} , 
\qquad k_y\in \frac{2\pi n}{W}
\end{equation}
with $n=0,\pm 1,\ldots$ being any integer.
In the mixed continuous-discrete $(k_x,k_y)$ Fourier space, the Coulomb
potential has the typical $1/k^2$ $(k^2=k_x^2+k_y^2)$ singularity as $k\to 0$. 
Integrating over the continuous variable $k_x$ and summing over the discrete
variable $k_y$, one obtains
\begin{eqnarray}
\phi(x,y) & = & - \ln \left\vert 2 \sinh\left( \frac{\pi z}{W} \right)
\right\vert \nonumber \\ & = & - \frac{1}{2}
\ln \left[ 2 \cosh\left( \frac{2\pi x}{W} \right) - 
2 \cos\left( \frac{2\pi y}{W} \right) \right]  \label{periodicpot}
\end{eqnarray}
with the complex number notation $z=x+{\rm i}y$ and $\bar{z}=x-{\rm i}y$.
For small distances $r\ll W$, this potential takes
the expected 2D logarithmic form $-\ln(2\pi r/W)$.
At large distances along the cylinder axis $x\gg W$, this potential
is the 1D Coulomb one $-\pi \vert x\vert/W$.

In what follows, the following integral for the periodic Coulomb potential
(\ref{periodicpot})
\begin{equation} 
\int_0^W {\rm d}y\, \phi({\bf r}) = - \frac{W}{4\pi} \int_0^{2\pi} {\rm d}y\,
\ln \left[ 2 \cosh\left( \frac{2\pi x}{W} \right) - 2 \cos y \right]
\end{equation}  
will be important.
Writing
\begin{equation} \label{defa}
2 \cosh\left( \frac{2\pi x}{W} \right) = a + \frac{1}{a} ,
\qquad a=\exp\left( \frac{2\pi x}{W} \right) \ge 1  
\end{equation}  
and using that
\begin{equation}
\int_0^{2\pi} {\rm d}y\, \ln \left( 1 + a^2 - 2 a \cos y \right)
= 4\pi \ln a , \qquad a\ge 1 ,
\end{equation}  
one obtains that the integral
\begin{equation} \label{usefulformula}
\int_0^W {\rm d}y\, \phi({\bf r}) = - \pi x 
\end{equation}
does not depend on the length $W$.
The important feature of this formula is the absence of an infinite
constant which occurs in the analogous integral over the pure 2D Coulomb
potential $-\frac{1}{2} \int_{-\infty}^{\infty} {\rm d}y\, \ln(x^2+y^2)$.

\subsection{Single EDL} \label{Section2.1}
\begin{figure}[t]
\begin{center}
\includegraphics[width=0.7\textwidth,clip]{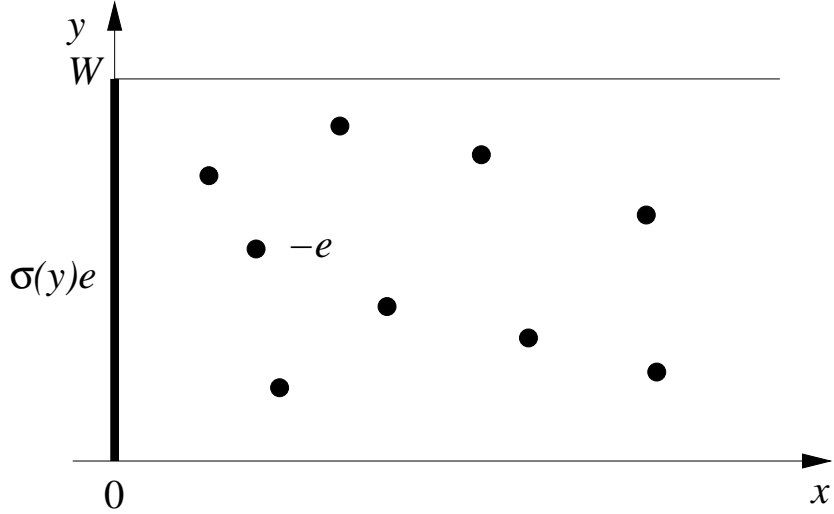}
\caption{The cylinder geometry with the periodic boundary conditions 
(period $W$) along the $y$-axis. 
The line (circle) at $x=0$ is charged by the line density $\sigma(y) e$.
Pointlike counterions of charge $-e$, represented by black circles,
move freely in the semi-infinite domain $\Lambda=\{ (x,y); x>0,y\in [0,W]\}$.}
\label{Fig:1}
\end{center}
\end{figure}

In the one-wall geometry of figure \ref{Fig:1}, there is an inhomogeneous
line charge density $\sigma(y) e$ ($y\in [0,W]$) fixed at $x=0$;
$\sigma(y)$ has the dimension of inverse length.
Being motivated by experiments with colloids, the sign of the line charge
density is supposed to be the same everywhere,
\begin{equation} \label{experiments}
\sigma(y)\ge 0 , \qquad y\in [0,W] .
\end{equation}  
The particles of the elementary charge $-e$ move in the semi-infinite space
$\Lambda=\{ (x,y); x>0,y\in [0,W]\}$.

The condition of the overall system electroneutrality reads as
\begin{equation}
N = \int_0^W {\rm d}y\, \sigma(y) .
\end{equation}
Under the effect of the inhomogeneous line charge density, the one-body
energy of the particle at position $(x,y)$ is given by
\begin{equation} \label{onebodyenergy1}
v(x,y) = - e^2 \int_0^W {\rm d}y'\, \sigma(y') \phi(x,y-y') . 
\end{equation}
Introducing the mean value of the line charge density
\begin{equation} \label{sigma}
\sigma \equiv \frac{1}{W} \int_0^W {\rm d}y\, \sigma(y) = \frac{N}{W}
\end{equation}
and its deviation from the mean value
\begin{equation}
\delta\sigma(y) \equiv \sigma(y)-\sigma , \qquad
\int_0^W {\rm d}y\, \delta\sigma(y) = 0 ,  
\end{equation}
using the integration formula (\ref{usefulformula}) one can write
the one-body energy (\ref{onebodyenergy1}) as follows
\begin{equation} \label{onebodyenergy2}
v(x,y) = \pi e^2 \sigma x
- e^2 \int_0^W {\rm d}y'\, \delta\sigma(y') \phi(x,y-y') . 
\end{equation}  

The Coulomb energy of $N$ particles at spatial positions
$\{ {\bf r}_1,\ldots,{\bf r}_N\}$ inside the domain $\Lambda$
plus the line charge density at $x=0$ consists of three parts,
$E_N=E_{ll}+E_{lp}+E_{pp}$, where
\begin{eqnarray} \label{Ell}
E_{ll} & = & \frac{e^2}{2} \int_0^W {\rm d}y \int_0^W {\rm d}y'
\sigma(y) \phi(0,y-y') \sigma(y')  \nonumber \\
& = & \frac{e^2}{2} \int_0^W {\rm d}y \int_0^W {\rm d}y'
\delta\sigma(y) \phi(0,y-y') \delta\sigma(y')  
\end{eqnarray}
is the self-energy of the line charge density at $x=0$,
\begin{equation} \label{Elp}
E_{lp} = \sum_{j=1}^N v(x_j,y_j) 
\end{equation}
describes the interaction of the line charge density with the particles and
\begin{equation} \label{Epp}
E_{pp} = e^2 \sum_{(j<k)=1}^N \phi({\bf r}_j-{\bf r}_k) =
- e^2 \sum_{(j<k)=1}^N \left\vert 2
\sinh\left( \frac{\pi (z_j-z_k)}{W} \right)\right\vert
\end{equation}
is the sum over all pair interactions among the particles.

The Boltzmann factor of the total energy at the inverse temperature
$\beta$ is given by
\begin{eqnarray}
{\rm e}^{-\beta E_N(\{{\bf r}\})} & = & {\rm e}^{-\frac{\Gamma}{2} \int_0^W {\rm d}y \int_0^W {\rm d}y'\delta\sigma(y)\phi(0,y-y') \delta\sigma(y')}
\prod_{j=1}^N {\rm e}^{-\beta v(x_j,y_j)} \nonumber \\ & & \times
\prod_{(j<k)=1}^N \left\vert 2 \sinh\frac{\pi(z_j-z_k)}{W} \right\vert^{\Gamma} ,
\label{Boltzmann}
\end{eqnarray}
where $\Gamma\equiv \beta e^2$ is the coupling constant.
Within the framework of the canonical ensemble, the partition function is
defined as
\begin{equation} \label{partition}
Z_N = \frac{1}{N!} \int_{\Lambda} \frac{{\rm d}{\bf r}_1}{\lambda^2} 
\cdots \int_{\Lambda} \frac{{\rm d}{\bf r}_N}{\lambda^2} 
{\rm e}^{-\beta E_N(\{{\bf r}\})} ,
\end{equation}
where $\lambda$ is the thermal de Broglie wavelength. 
Applying the formula
\begin{equation} \label{formula2}
\left\vert 2 \sinh \frac{\pi (z-z')}{W} \right\vert
= {\rm e}^{\frac{\pi}{W}(x+x')} \left\vert
{\rm e}^{-\frac{2\pi}{W}z} - {\rm e}^{-\frac{2\pi}{W}z'} \right\vert
\end{equation}
for each two-body Boltzmann factor, the partition function (\ref{partition}) 
is expressible as
\begin{equation} \label{partf}
Z_N = \left( \frac{W^2}{4\pi\lambda^2} \right)^N 
{\rm e}^{-\frac{\Gamma}{2} \int_0^W {\rm d}y \int_0^W {\rm d}y'\, \delta\sigma(y) \phi(0,y-y') \delta\sigma(y')}
Q_N ,
\end{equation}
where
\begin{equation} \label{partgen}
Q_N = \frac{1}{N!} \int_{\Lambda} \prod_{j=1}^N \left[ {\rm d}^2 z_j\,
w({\bf r}_j) \right] \prod_{(j<k)=1}^N \left\vert
{\rm e}^{-\frac{2\pi}{W}z_j} - {\rm e}^{-\frac{2\pi}{W}z_k} \right\vert^{\Gamma} 
\end{equation}
with the renormalized one-body weight function
\begin{equation} \label{onebody1}
w({\bf r}) = \frac{4\pi}{W^2} \exp\left[ -\beta v(x,y)
+ \frac{\pi\Gamma}{W} (N-1) x \right] .
\end{equation}
With regard to the equality (\ref{onebodyenergy2}), one has
\begin{equation} \label{onebody2}
w({\bf r}) = \frac{4\pi}{W^2} \exp\left[ - \frac{\pi\Gamma}{W} x
+\Gamma \int_0^W {\rm d}y'\, \delta\sigma(y') \phi(x,y-y') \right] .
\end{equation}
The free energy $F_N$ is defined by $-\beta F_N = \ln Z_N$.
With respect to (\ref{partf}), it is expressible as
\begin{eqnarray} 
-\beta F_N(\gamma) & = & N \ln \left( \frac{W^2}{4\pi\lambda^2} \right) 
- \frac{\Gamma}{2} \int_0^W {\rm d}y \int_0^W {\rm d}y'\,
\delta\sigma(y) \phi(0,y-y') \delta\sigma(y') \nonumber \\ & &
+ \ln Q_N(\gamma) . \label{free1}
\end{eqnarray}

Introducing the microscopic particle number density
$\hat{n}({\bf r}) = \sum_{j=1}^N \delta({\bf r}-{\bf r}_j)$,
the mean particle number density at point ${\bf r}\in \Lambda$ is given by 
\begin{equation}
n({\bf r}) = \left\langle \hat{n}({\bf r}) \right\rangle , 
\qquad 
\end{equation}
where $\langle \cdots \rangle$ denotes the statistical average over
the canonical ensemble.
Since in the partition function (\ref{partf}) only the multiple integral $Q_N$
involves particle coordinates, the particle density can be obtained as
the functional derivative
\begin{equation} \label{onefunct}
n({\bf r}) = w({\bf r}) \frac{1}{Q_N} 
\frac{\delta Q_N}{\delta w({\bf r})} .
\end{equation}
Since there is a finite number of particles per unit length of the line wall,
the particle number density must vanish at large distances from the wall,
\begin{equation} \label{bc}
\lim_{x\to\infty} n(x,y) = 0 .
\end{equation}  

\subsection{Two parallel EDLs} \label{Section2.2}
\begin{figure}[t]
\begin{center}
\includegraphics[width=0.7\textwidth,clip]{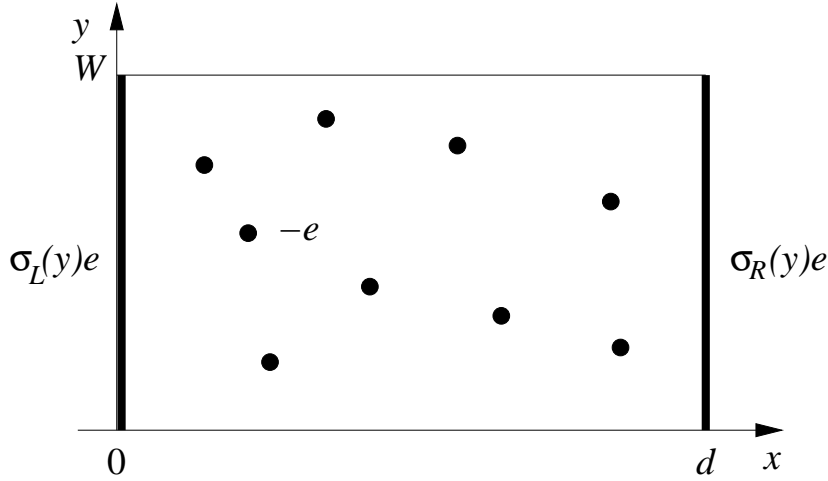}
\caption{The cylinder geometry with the periodic boundary conditions 
(period $W$) along the $y$-axis. 
Two parallel lines (circles) with the fixed charge densities $\sigma_L(y) e$ 
and $\sigma_R(y) e$ are localized at the left and right end points
$x=0$ and $x=d$, respectively. 
Pointlike counterions of charge $-e$ (black circles) move freely between 
the two charged lines.}
\label{Fig:2}
\end{center}
\end{figure}

In the case of two parallel walls at distance $d$ (see figure \ref{Fig:2}),
there are ``left'' and ``right'' line charge densities
$\sigma_L(y) e$ and $\sigma_R(y) e$ fixed on the line segments at
the end-points $x=0$ and $x=d$, respectively, such that
\begin{equation}
\sigma_L(y)\ge 0 , \quad \sigma_R(y)\ge 0 , \qquad y\in [0,W] .
\end{equation}  
The mobile charges are constrained to the space between the walls
$\Lambda=\{ (x,y); 0<x<d,y\in [0,W]\}$.
The electroneutrality condition reads as
\begin{equation}
N = \int_0^W {\rm d}y\, \left[ \sigma_L(y) + \sigma_R(y) \right] .
\end{equation} 
The mean values of the left and right line charge densities
\begin{equation}
\sigma_L \equiv \frac{1}{W} \int_0^W {\rm d}y\, \sigma_L(y) , \qquad
\sigma_R \equiv \frac{1}{W} \int_0^W {\rm d}y\, \sigma_R(y)   
\end{equation}
are constrained by
\begin{equation} \label{LLRR}
\sigma_L + \sigma_R = \frac{N}{W} .
\end{equation}
The deviations of the left and right line charge densities
from their mean values are defined by
\begin{equation}
\delta\sigma_L(y) \equiv \sigma_L(y)-\sigma_L , \qquad
\int_0^W {\rm d}y\, \delta\sigma_L(y) = 0   
\end{equation}
and
\begin{equation}
\delta\sigma_R(y) \equiv \sigma_R(y)-\sigma_R , \qquad
\int_0^W {\rm d}y\, \delta\sigma_R(y) = 0 ,
\end{equation}
respectively.
The one-body energy of the particle at position $(x,y)$ reads as
\begin{eqnarray} 
v(x,y) & = & - e^2 \int_0^W {\rm d}y'\, \sigma_L(y') \phi(x,y-y')
\nonumber \\ & &
- e^2 \int_0^W {\rm d}y'\, \sigma_R(y') \phi(d-x,y-y') . \label{onebodyenergy3} 
\end{eqnarray}
Using (\ref{usefulformula}), it can be written as
\begin{eqnarray} 
v(x,y) & = & \pi e^2 \sigma_R d + \pi e^2 \left( \sigma_L - \sigma_R \right) x
- e^2 \int_0^W {\rm d}y'\, \delta\sigma_L(y') \phi(x,y-y') \nonumber \\ 
& & - e^2 \int_0^W {\rm d}y'\, \delta\sigma_R(y') \phi(d-x,y-y') . 
\label{onebodyenergy4}
\end{eqnarray}  

The total Coulomb energy of $N$ particles at spatial positions
$\{ {\bf r}_1,\ldots,{\bf r}_N\}$ inside the domain $\Lambda$ plus the
left and right line charge densities at $x=0$ and $x=d$, respectively,
consists of three parts: $E_N = E_{ll} + E_{lp} + E_{pp}$.
The self-energies and the interaction energy of the line charge densities
at $x=0$ and $x=d$ are given by
\begin{eqnarray}
E_{ll} & = & \frac{e^2}{2} \int_0^W {\rm d}y \int_0^W {\rm d}y'
\sigma_L(y) \phi(0,y-y') \sigma_L(y') \nonumber \\
&  & + \frac{e^2}{2} \int_0^W {\rm d}y \int_0^W {\rm d}y'
\sigma_R(y) \phi(0,y-y') \sigma_R(y') \nonumber \\
& & +  e^2 \int_0^W {\rm d}y \int_0^W {\rm d}y'
\sigma_L(y) \phi(d,y-y') \sigma_R(y') \nonumber \\
& = & \frac{e^2}{2} \int_0^W {\rm d}y \int_0^W {\rm d}y'
\delta\sigma_L(y) \phi(0,y-y') \delta\sigma_L(y') \nonumber \\
&  & + \frac{e^2}{2} \int_0^W {\rm d}y \int_0^W {\rm d}y'
\delta\sigma_R(y) \phi(0,y-y') \delta\sigma_R(y') \nonumber \\
& & +  e^2 \int_0^W {\rm d}y \int_0^W {\rm d}y'
\delta\sigma_L(y) \phi(d,y-y') \delta\sigma_R(y') \nonumber \\
& & -\pi e^2 \sigma_L \sigma_R W d .
\end{eqnarray}
The interaction energy of the line charge densities with the particles
$E_{lp}$ is given by formula (\ref{Elp}) with the one-body potential $v(x,y)$
substituted from (\ref{onebodyenergy4}).
The formula (\ref{Epp}) for the particle-particle interaction
$E_{pp}$ remains unchanged.

Repeating the procedure in the paragraph between equations (\ref{Boltzmann})
and (\ref{free1}) accomplished for the one-wall case, the (dimensionless)
free energy takes the form
\begin{eqnarray} 
-\beta F_N & = & N \ln \left( \frac{W^2}{4\pi\lambda^2} \right) 
- \frac{\Gamma}{2} \int_0^W {\rm d}y \int_0^W {\rm d}y'\,
\delta\sigma_L(y) \phi(0,y-y') \delta\sigma_L(y') \nonumber \\ & &
- \frac{\Gamma}{2} \int_0^W {\rm d}y \int_0^W {\rm d}y'\,
\delta\sigma_R(y) \phi(0,y-y') \delta\sigma_R(y') \nonumber \\ & &
- \Gamma \int_0^W {\rm d}y \int_0^W {\rm d}y'\,
\delta\sigma_L(y) \phi(d,y-y') \delta\sigma_R(y') \nonumber \\ & &
-\pi \Gamma \sigma_R^2 W d + \ln Q_N , 
\label{free2}
\end{eqnarray}
where $Q_N$ is defined by (\ref{partgen}) with the renormalized
one-body weight function (\ref{onebody1}) of the form 
\begin{eqnarray} 
w(x,y) & = & \frac{4\pi}{W^2} \exp\Bigg\{ - \frac{\pi\Gamma}{W} x
+ 2 \pi \Gamma \sigma_R x + \Gamma \int_0^W {\rm d}y'\,
\delta\sigma_L(y') \phi(x,y-y') \nonumber \\ & &
+ \Gamma \int_0^W {\rm d}y'\, \delta\sigma_R(y') \phi(d-x,y-y') \Bigg\} .
\label{onebody3}
\end{eqnarray}
The pressure $P_N$, i.e. the force between the left and right charged line
fragments (per unit length of one of the lines), is given by
\begin{eqnarray}
\beta P_N & = & \frac{\partial}{\partial d}
\left( \frac{-\beta F_N}{W} \right) \nonumber \\
& = & -\frac{\Gamma}{W} \int_0^W {\rm d}y \int_0^W {\rm d}y'\,
\delta\sigma_L(y) \frac{\partial\phi(d,y-y')}{\partial d} \delta\sigma_R(y') 
\nonumber \\ & &
- \pi\Gamma\sigma_R^2 + \frac{1}{W Q_N} \frac{\partial Q_N}{\partial d} .
\label{pressure}
\end{eqnarray}  

The calculation of the one-body density by using the generating functional
$Q_N$ proceeds as in the one-wall case, see equation (\ref{onefunct}).

\renewcommand{\theequation}{3.\arabic{equation}}
\setcounter{equation}{0}

\section{Formalism of anticommuting variables} \label{Section3}

\subsection{Mapping onto the 1D lattice anticommuting-field theory}
\label{Section3.1}
For $\Gamma=2\gamma$ ($\gamma=1,2,3,\ldots$ a positive integer),
the generating functional $Q_N$ (\ref{partgen}) can be expressed
as an integral over Grassmann (anticommuting) variables
\cite{Samaj95,Samaj04a}; for the cylinder geometry the mapping
is established in \cite{Samaj04b,Samaj14}.

Let us consider a discrete chain of $N$ sites $j=0,1,\ldots,N-1$.
At each site $j$, there is $\gamma$ variables of type $\{ \xi_j^{(\alpha)}\}$
and $\gamma$ variables of type $\{ \psi_j^{(\alpha)}\}$
$(\alpha=1,\ldots,\gamma)$, the variables anticommute with each other. 
The multi-dimensional integral of the form (\ref{partgen}) is expressible as
the integral over anticommuting variables as follows
\begin{equation} \label{antipart}
Q_N = \int {\cal D}\psi {\cal D}\xi\, {\rm e}^{S(\Xi,\Psi)} , 
\qquad S(\Xi,\Psi) = \sum_{j,k=0}^{\gamma(N-1)} \Xi_j w_{jk} \Psi_k .
\end{equation}
Here, ${\cal D}\psi {\cal D}\xi \equiv \prod_{j=0}^{N-1} {\rm d}\psi_j^{(\gamma)}
\cdots {\rm d}\psi_j^{(1)} {\rm d}\xi_j^{(\gamma)} \cdots {\rm d}\xi_j^{(1)}$
and the action $S(\Xi,\Psi)$ involves pair interactions of composite
operators
\begin{equation} \label{composite}
\Xi_j = \sum_{j_1,\ldots,j_{\gamma}=0\atop (j_1+\cdots+j_{\gamma}=j)}^{N-1}
\xi_{j_1}^{(1)} \cdots \xi_{j_{\gamma}}^{(\gamma)} , \qquad
\Psi_j = \sum_{j_1,\ldots,j_{\gamma}=0\atop (j_1+\cdots+j_{\gamma}=j)}^{N-1}
\psi_{j_1}^{(1)} \cdots \psi_{j_{\gamma}}^{(\gamma)} ,
\end{equation} 
i.e. the products of $\gamma$ anticommuting variables of one type with 
the prescribed sum of site indices.
The interaction strengths $w_{jk}$ $[j,k=0,1,\ldots,\gamma(N-1)]$ are given by
\begin{equation} \label{wjk}
w_{jk} = \int_{\Lambda} {\rm d}^2z\, w(z,\bar{z}) 
\exp\left( - \frac{2\pi}{W} j z \right)
\exp\left( - \frac{2\pi}{W} k \bar{z} \right) .
\end{equation}

The particle number density at the point ${\bf r}=(z,\bar{z})$ is given by
\begin{equation} \label{antione}
n({\bf r}) = w({\bf r}) \sum_{j,k=0}^{\gamma(N-1)}
\langle \Xi_j \Psi_k \rangle
\exp\left( - \frac{2\pi}{W} j z \right)
\exp\left( - \frac{2\pi}{W} k \bar{z} \right) ,
\end{equation}
where
\begin{equation}
\langle \cdots\rangle \equiv \frac{1}{Q_N}
\int {\cal D}\psi {\cal D}\xi\, {\rm e}^{S(\Xi,\Psi)} \cdots .
\end{equation}
denotes the average over the anticommuting variables.

\subsection{Exactly solvable coupling $\Gamma=2$} \label{Section3.2}
The formalism simplifies itself for the coupling $\Gamma=2$ ($\gamma=1$)
when the composite operators $\{ \Xi_j,\Psi_j\}_{j=0}^{N-1}$ become the standard
anticommuting variables $\{ \xi_j,\psi_j\}_{j=0}^{N-1}$.
Due to the bilinear form of the action $S=\sum_{j,k=0}^{N-1}\xi_j w_{jk} \psi_k$,
it holds that
\begin{eqnarray}
Q_N & = & {\rm Det}(w_{jk})\vert_{j,k=0}^{N-1} ,  \\
\langle \xi_j \psi_k \rangle & = & w^{-1}_{kj} .   
\end{eqnarray}

The $N\times N$ matrix $w_{jk}$ can be inverted only in specific cases.
The most trivial case is the diagonal form of the matrix
$w_{jk} = w_j \delta_{jk}$ $(j,k=0,1,\ldots,N-1)$ when
\begin{equation}
Q_N = \prod_{j=0}^{N-1} w_j , \qquad
w^{-1}_{kj} = w^{-1}_{jk} = \frac{1}{w_j} \delta_{jk} .  
\end{equation}
Thus, the particle number density is given by
\begin{equation} \label{nxy}
n(x,y) = w(x,y) \sum_{j=0}^{N-1} \frac{1}{w_j}
\exp\left( - \frac{4\pi}{W} j x\right) . 
\end{equation}

\subsection{Sum rules} \label{Section3.3}
There exist linear transformations of anticommuting variables
$\{ \xi \}$ or $\{ \psi \}$ which maintain the composite form of
the operators $\{ \Xi \}$ or $\{ \Psi \}$ \cite{Samaj00}.
Every transformation implies exact constraints (sum rules) for the correlators
$\langle \Xi_j \Psi_k \rangle$ and, consequently,
integral equations for one-body densities.
Here, we make use of two simplest transformations.

$\bullet$ Let at each site one of the field components, say $\{ \xi^{(1)} \}$,
be rescaled by a constant $\mu$:
\begin{equation} \label{trans1}
\xi_j^{(1)} \to \mu \xi_j^{(1)} , \qquad j=0,1,\ldots,N-1 .
\end{equation}
The Jacobian of this transformation equals to $\mu^N$.
The composite operators $\{ \Xi \}$ get the same factor $\mu$ and the action
in (\ref{antipart}) is transformed as follows
$S(\Xi,\Psi)\to \mu S(\Xi,\Psi)$.
Consequently,
\begin{equation}
Q_N = \mu^{-N} \int {\cal D}\psi {\cal D}\xi
\exp\left( \mu \sum_{j,k=0}^{\gamma(N-1)} \Xi_j w_{jk} \Psi_k \right) .
\end{equation}

Since $Q_N$ is independent of $\mu$, its derivative with respect to $\mu$
is zero for any value of $\mu$.
Choosing $\mu=1$, the equality $\partial \ln Q_N/\partial\mu\vert_{\mu=1} = 0$
is equivalent to
\begin{equation} \label{sr11}
\sum_{j,k=0}^{\gamma(N-1)} w_{jk} \langle \Xi_j \Psi_k \rangle = N .
\end{equation}
Using the definition of the interaction strength $w_{jk}$ (\ref{wjk})
and the representation (\ref{antione}) of $n(z,\bar{z})$,
this relation is equivalent to the trivial information 
\begin{equation} \label{ie11}
\int_{\Lambda} {\rm d}^2z\, n(z,\bar{z}) = N 
\end{equation}
that there are just $N$ particles in the cylinder domain $\Lambda$.

$\bullet$ Let us rescale at each site {\em all} anticommuting field
$\xi$-components as follows
\begin{equation} \label{trans21}
\xi_j^{(\alpha)} \to \lambda^j \xi_j^{(\alpha)} , \qquad \alpha=1,\ldots,\gamma,
\qquad j=0,1,\ldots,N-1,
\end{equation}
or {\em all} anticommuting field $\psi$-components,
\begin{equation} \label{trans22}
\psi_j^{(\alpha)} \to \lambda^j \psi_j^{(\alpha)} , \qquad \alpha=1,\ldots,\gamma,
\qquad j=0,1,\ldots,N-1.
\end{equation}
The Jacobian of both transformations equals to $\lambda^{\gamma N(N-1)/2}$. 
Under the transformation (\ref{trans21}), the composite operators $\Xi_j$
acquire the factor $\lambda^j$ and the action in (\ref{antipart}) transforms
itself to $S(\Xi,\Psi)\to \sum_{j,k=0}^{\gamma(N-1)} \lambda^j \Xi_j w_{jk} \Psi_k$.
Thus,
\begin{equation}
Q_N = \lambda^{-\gamma N(N-1)/2} \int {\cal D}\psi {\cal D}\xi
\exp\left( \sum_{j,k=0}^{\gamma(N-1)} \lambda^j \Xi_j w_{jk} \Psi_k \right) .
\end{equation}
Analogously, under the transformation (\ref{trans22}), the composite
operators $\Psi_j$ acquire the factor $\lambda^j$ and the action in
(\ref{antipart}) transforms itself to
$S(\Xi,\Psi)\to \sum_{j,k=0}^{\gamma(N-1)} \lambda^k \Xi_j w_{jk} \Psi_k$,
i.e.,
\begin{equation}
Q_N = \lambda^{-\gamma N(N-1)/2} \int {\cal D}\psi {\cal D}\xi
\exp\left( \sum_{j,k=0}^{\gamma(N-1)} \lambda^k \Xi_j w_{jk} \Psi_k \right) .
\end{equation}

The requirement $\partial \ln Q_N/\partial\lambda\vert_{\lambda=1} = 0$ 
is equivalent to the sum rules
\begin{eqnarray} 
\sum_{j,k=0}^{\gamma(N-1)} j w_{jk} \langle \Xi_j \Psi_k \rangle 
& = & \frac{1}{2}\gamma N (N-1) , \label{sr21} \\
\sum_{j,k=0}^{\gamma(N-1)} k w_{jk} \langle \Xi_j \Psi_k \rangle 
& = & \frac{1}{2}\gamma N (N-1) . \label{sr22}
\end{eqnarray}
These equalities can be rewritten into a more suitable form
\begin{eqnarray} 
\sum_{j,k=0}^{\gamma(N-1)} (j+k) w_{jk} \langle \Xi_j \Psi_k \rangle 
& = & \gamma N (N-1) , \label{sr21prime} \\
\sum_{j,k=0}^{\gamma(N-1)} (j-k) w_{jk} \langle \Xi_j \Psi_k \rangle 
& = & 0 . \label{sr22prime}
\end{eqnarray}

\renewcommand{\theequation}{4.\arabic{equation}}
\setcounter{equation}{0}

\section{Analytic results for one EDL} \label{Section4}

\subsection{Derivation of the contact value theorem} \label{Section4.1}
With regard to the definition of the interaction strengths (\ref{wjk}),
the first sum rule (\ref{sr21prime}) can be expressed as
\begin{eqnarray}
\int_{\Lambda} {\rm d}^2z\, w(z,\bar{z}) \sum_{j,k=0}^{\gamma(N-1)}
(j+k) \langle \Xi_j\Psi_k \rangle
\exp\left[ -\frac{2\pi}{W} (j+k)x -\frac{2\pi}{W} (j-k) {\rm i}y \right]
\nonumber \\ = \gamma N(N-1) .  
\end{eqnarray}
Using the expression for the particle density (\ref{antione}),
this equation is expressed as
\begin{equation}
- \frac{W}{2\pi} \int_{\Lambda} {\rm d}^2z\, w(z,\bar{z})
\frac{\partial}{\partial x} \left[ \frac{n(z,\bar{z})}{w(z,\bar{z})} \right]
= \frac{\Gamma}{2} N (N-1) .  
\end{equation}  
Thus,
\begin{eqnarray}
- \int_{\Lambda} {\rm d}^2r\, \frac{\partial}{\partial x} n(x,y)
+ \int_{\Lambda} {\rm d}^2r\, n(x,y)
\frac{\partial}{\partial x} \left[ \ln w(x,y) \right] 
& = & \frac{\pi\Gamma}{W} N (N-1) . \nonumber \\ & &  \label{crucialeq}
\end{eqnarray}  
For the geometry of one wall with the one-body weight function
$w(x,y)$ (\ref{onebody2}), it holds that
\begin{equation}
\frac{\partial}{\partial x} \ln w(x,y) = - \frac{\pi\Gamma}{W}
+ \Gamma \int_0^W {\rm d}y'\, \delta\sigma(y')
\frac{\partial \phi(x,y-y')}{\partial x} .
\end{equation}  
Taking into account the boundary condition (\ref{bc}) and the equality
(\ref{sigma}), one finally gets the contact value theorem
\begin{eqnarray} 
\int_0^W {\rm d}y\,  \left[ n(0,y) - \pi \Gamma \sigma^2 \right]
+ \Gamma \int_0^W {\rm d}y \int_0^{\infty} {\rm d}x \, n(x,y)
\nonumber \\ \times \int_0^W {\rm d}y'\, \delta\sigma(y') 
\frac{\partial \phi(x,y-y')}{\partial x} = 0 . \label{cvt}
\end{eqnarray}
Although the outlined derivation was performed only for the discrete series of
the coupling constants $\Gamma/2=$integer, it is reasonable to anticipate
its validity also for continuous values of $\Gamma$.

In the case of a homogeneously charged line with $\delta\sigma(y)=0$,
this relation reduces itself to the standard 2D contact value theorem
relating the contact density $n(0,y) = n(0)$ to the uniform
line charge density $\sigma$ as follows
\begin{equation} \label{cvtheorem}
n(0)=\Gamma\pi\sigma^2 .
\end{equation}  
Note that in the presence of the line charge modulation the contact
density $n(0,y)$, integrated over the domain boundary, depends on
the density profile $n(x,y)$ inside the whole region $\Lambda$.

The second sum rule (\ref{sr22prime}) is equivalent to the condition
\begin{equation}
\int_{\Lambda} {\rm d}^2z\, w(z,\bar{z}) \frac{\partial}{\partial y}
\left[ \frac{n(z,\bar{z})}{w(z,\bar{z})} \right] = 0.  
\end{equation}
Taking into account the continuity and periodicity of the density along
the $y$-axis, $n(x,W)=n(x,0)$, this condition is equivalent to the one
\begin{equation}
\int_0^W {\rm d}y \int_0^{\infty} {\rm d}x\, n(x,y)
\int_0^W {\rm d}y'\, \delta\sigma(y')
\frac{\partial \phi(x,y-y')}{\partial y} = 0 .
\end{equation}

\subsection{The exactly solvable coupling $\Gamma=2$} \label{Section4.2}
For the free-fermion coupling $\Gamma=2$, the renormalized one-body weight
function (\ref{onebody2}) takes the form
\begin{equation} \label{renorm1}
w(x,y) = \frac{4\pi}{W^2} \exp\left[ - \frac{2\pi}{W} x
+ 2 \int_0^W {\rm d}y'\, \delta\sigma(y') \phi(x,y-y') \right] .
\end{equation}
We assume that the deviation of the line charge density from its
mean value is a continuous and periodic function of the $y$-coordinate,
\begin{equation} \label{period}
\delta\sigma(y) = \delta\sigma(y+\lambda) , \qquad
\lambda = \frac{W}{M} \quad (M=1,2,\ldots) . 
\end{equation}  
Note that the condition for $M$ being a positive integer ensures
the continuity of $\delta\sigma(y)$ across the cylinder boundary along the
$y$-axis, $\delta\sigma(y) = \delta\sigma(y+W)$.
The periodicity of $\delta\sigma(y)$ automatically means
the periodicity of the one-body weight function (\ref{renorm1}),
\begin{equation} \label{renorm2}
w(x,y) = w(x,y+\lambda) .
\end{equation}
Within the anticommuting-field representation, the interaction
strengths (\ref{wjk}) are given by
\begin{eqnarray}
w_{jk} & = & \int_0^{\infty} {\rm d}x\, {\rm e}^{-\frac{2\pi}{W}(j+k)x}
U(j-k,x) , \nonumber \\
U(\alpha,x) & = & \int_0^W {\rm d}y\, w(x,y)
{\rm e}^{-{\rm i}\frac{2\pi}{W}\alpha y} .
\end{eqnarray}
In the diagonal case $j=k$, the integral
\begin{equation}
U(0,x) = M \int_0^{\lambda} {\rm d}y\, w(x,y) > 0
\end{equation}
and so the diagonal elements
\begin{equation}
w_{jj} \equiv w_j = \int_0^{\infty} {\rm d}x\, {\rm e}^{-\frac{4\pi}{W}jx} U(0,x) 
\end{equation}
are positive numbers.
For the off-diagonal $j\ne k$ elements, the integral $U(\alpha,x)$
with $\alpha=\pm 1,\pm 2,\ldots,\pm (N-1)$ can be expressed as
\begin{equation} \label{geomseries}
U(\alpha,x) = \int_0^{\lambda} {\rm d}y\, w(x,y) 
{\rm e}^{-{\rm i}\frac{2\pi}{W}\alpha y} \left( 1 + q + q^2 + \cdots
+ q^{M-1} \right) ,
\end{equation}
where $q=\exp\left( -{\rm i}\frac{2\pi}{W}\alpha\lambda \right)=
\exp\left( -{\rm i}\frac{2\pi}{M}\alpha\right)$.
To ensure that $q\ne 1$, let us assume that $M\ge N$ or, with regard to
the equalities $M=W/\lambda$ and $N=\sigma W$,
\begin{equation} \label{cond}
\lambda \sigma \le 1 .
\end{equation}  
Note that the equality $\lambda \sigma=1$ corresponds to the most interesting
case when there is just one counterion per the surface-charge period
(i.e. there is one electron released by one ``blurred'' nucleus at the wall).
The summation of the finite geometric series in (\ref{geomseries})
then results in
\begin{equation}
\frac{q^M-1}{q-1} = \frac{{\rm e}^{-{\rm i}2\pi\alpha}-1}{q-1} = 0 .
\end{equation}  
This means that the matrix of interaction strengths is diagonal,
$w_{jk}=w_j \delta_{jk}$, with the positive diagonal elements
\begin{equation} \label{diagel}
w_j = \frac{W}{\lambda} \int_0^{\infty} {\rm d}x\,
\exp\left( -\frac{4\pi}{W} j x \right)
\int_0^{\lambda} {\rm d}y\, w(x,y) ,
\end{equation}
$j=0,1,\ldots,N-1$.
The particle number density is given by the relation (\ref{nxy}).
In the case of the opposite inequality $M<N$, the $N\times N$ matrix of
interaction strengths $w_{jk}$ is no longer diagonal and as such cannot be
inverted analytically; the model is thus not explicitly solvable. 

The previous analysis was general, valid for any function $\delta\sigma(y)$
satisfying the periodicity condition (\ref{period}).  
For simplicity, let us now restrict ourselves on the periodic cosine
dependence
\begin{equation} \label{deltasigma}
\delta\sigma(y)  = A \cos\left( \frac{2\pi}{\lambda} y \right) .  
\end{equation}
This form of the surface charge modulation describes adequately real solid
electrodes consisting of a periodic array of positively charged heavy nuclei,
bound together with core electrons and vibrating thermally around their
local equilibrium positions, and delocalized (in the case of conductors)
light valence electrons. 
To ensure the physical requirement (\ref{experiments}), the real amplitude
$A$ with dimension of inverse length must obey the inequality
\begin{equation} \label{inequality}
\vert A \vert \le \sigma .
\end{equation}
The integral term in the exponential of the definition of $w(x,y)$
(\ref{renorm1})
\begin{equation} \label{Vxy}
V(x,y) = 2 \int_0^W {\rm d}y'\, \delta\sigma(y') \phi(x,y-y')
\end{equation}
can be expressed as
\begin{eqnarray} 
V(x,y) & = & -A \int_0^W {\rm d}y'\, \cos\left( \frac{2\pi}{\lambda} y' \right)
\ln \left\{ 1 + a^2 - 2 a \cos\left[ \frac{2\pi}{W} (y-y') \right]
\right\} , \nonumber \\ & & \label{int1}
\end{eqnarray}
where the parameter $a=\exp\left( 2\pi x/W\right)\ge 1$ has already been
introduced in (\ref{defa}).
Using the substitution $t=2\pi(y'-y)/W$ in the integral (\ref{int1})
implies
\begin{equation} \label{int2}
V(x,y) = -A \frac{W}{2\pi} \int_0^{2\pi} {\rm d}t\,
\cos\left( \frac{2\pi}{\lambda} y + M t \right)
\ln \left( 1 + a^2 - 2 a \cos t \right) .
\end{equation}
Writing
$\cos\left( \frac{2\pi}{\lambda} y + M t \right) = \cos\left(
\frac{2\pi}{\lambda} y\right) \cos(M t) -
\sin\left( \frac{2\pi}{\lambda} y\right) \sin(M t)$
and using the equalities
\begin{eqnarray}
\int_0^{2\pi} {\rm d}t\, \sin(M t) \ln \left( 1 + a^2 - 2 a \cos t \right)
& = & 0 , \\  
\int_0^{2\pi} {\rm d}t\, \cos(M t) \ln \left( 1 + a^2 - 2 a \cos t \right)
& = & - \frac{2\pi}{M} a^{-M}     
\end{eqnarray}  
valid for $M=1,2,\ldots$ and $a>1$, one obtains that
\begin{equation} \label{Vxyresult}
V(x,y) = \lambda A \exp\left( - \frac{2\pi}{\lambda} x \right)
\cos\left( \frac{2\pi}{\lambda} y \right) .  
\end{equation}  
Finally, since
\begin{equation}
w(x,y) = \frac{4\pi}{W^2} \exp\left[ - \frac{2\pi}{W} x
+ \lambda A {\rm e}^{-\frac{2\pi}{\lambda}x}
\cos\left( \frac{2\pi}{\lambda} y\right) \right] ,   
\end{equation}  
the diagonal elements (\ref{diagel}) take the form
\begin{equation}
w_j = \frac{4\pi}{W} \int_0^{\infty} {\rm d}x\,
{\rm e}^{-\frac{4\pi}{W}(j+\frac{1}{2})x}
I_0\left( \lambda A {\rm e}^{-\frac{2\pi}{\lambda}x} \right) ,
\end{equation}
where
\begin{equation} \label{Bessel}
I_0(t) = \frac{1}{\pi} \int_0^{\pi} {\rm d}\varphi\, {\rm e}^{t\cos\varphi}   
\end{equation}  
is the Bessel function of the first kind with imaginary argument
\cite{Gradshteyn}.

According to the formula (\ref{nxy}), the particle number density is yielded by
\begin{equation}
n(x,y) = \frac{4\pi}{W^2} \exp\left[ \lambda A {\rm e}^{-\frac{2\pi}{\lambda}x}
\cos\left( \frac{2\pi}{\lambda} y\right) \right] \sum_{j=0}^{N-1} \frac{1}{w_j}
\exp^{-\frac{4\pi}{W} \left( j+\frac{1}{2} \right) x} .
\end{equation}
The sum over $j$ simplifies itself in the thermodynamic limit
$N,W\to\infty$, with the fixed ratio $N/W = \sigma$; in this limit,
the wall corresponds to an infinite line and charges interact via
the pure 2D logarithmic Coulomb potential.
Introducing the continuous variable
$t=\frac{1}{N}\left( j+\frac{1}{2} \right)$ and
transforming the sum $\frac{1}{N} \sum_{j=0}^{N-1}$ to the
integral $\int_0^1 {\rm d}t$, one arrives at
\begin{eqnarray}
n(x,y) & = & \frac{N}{W} \exp\left[ \lambda A {\rm e}^{-\frac{2\pi}{\lambda}x}
\cos\left( \frac{2\pi}{\lambda} y\right) \right] \nonumber \\
& & \times \int_0^1 {\rm d}t\,
\frac{{\rm e}^{-4\pi\sigma t x}}{\int_0^{\infty} {\rm d}x'\,
{\rm e}^{-4\pi\sigma t x'}
I_0\left( \lambda A {\rm e}^{-\frac{2\pi}{\lambda}x'} \right)} .
\label{nxyalmostfinal}
\end{eqnarray}
Finally, with the aid of the substitution $s=4\pi\sigma t$,
one gets
\begin{eqnarray}
n(x,y) & = & \frac{1}{4\pi} \exp\left[ \lambda A {\rm e}^{-\frac{2\pi}{\lambda}x}
\cos\left( \frac{2\pi}{\lambda} y\right) \right] \nonumber \\
& & \times \int_0^{4\pi\sigma} {\rm d}s\,
\frac{{\rm e}^{-s x}}{\int_0^{\infty} {\rm d}x'\,
{\rm e}^{-s x'} I_0\left( \lambda A {\rm e}^{-\frac{2\pi}{\lambda}x'} \right)} .
\label{nxyfinal}
\end{eqnarray}  
This particle number density is periodic along the $y$-axis,
with the same period $\lambda$ as the line charge density.
It must obey, within the interval of one period, 
the electroneutrality condition
\begin{equation}
\frac{1}{\lambda} \int_0^{\infty} {\rm d}x \int_0^{\lambda} {\rm d}y\,
n(x,y) = \sigma .
\end{equation}
It is easy to check that our result (\ref{nxyfinal}) indeed satisfies this
constraint.

Based on the proven equivalence of equations (\ref{Vxy}) and (\ref{Vxyresult}),
it holds that
\begin{equation}
\frac{\partial}{\partial x} \int_0^W {\rm d}y'\, \delta\sigma(y')
\phi(x,y-y') = - \pi A \exp\left( - \frac{2\pi}{\lambda} x \right)
\cos\left( \frac{2\pi}{\lambda} y \right) .    
\end{equation}  
Considering this equation in the contact value theorem (\ref{cvt})
taken at $\Gamma=2$ and reducing the integration interval over $y\in [0,W]$
to a series of equivalent integrals over one period $\lambda$, one obtains
a simplified form of the contact value theorem
\begin{eqnarray} 
\int_0^{\lambda} {\rm d}y\,  \left[ n(0,y) - 2 \pi \sigma^2 \right]
= 2\pi A \int_0^{\lambda} {\rm d}y \int_0^{\infty} {\rm d}x \, n(x,y)
{\rm e}^{- \frac{2\pi}{\lambda} x} \cos\left( \frac{2\pi}{\lambda} y \right) .
\nonumber \\ \label{cvtsimple}
\end{eqnarray}  
Inserting here the result for the counterion density (\ref{nxyfinal})
and using that
\begin{eqnarray}
- 2 \pi A {\rm e}^{- \frac{2\pi}{\lambda} x}
\cos\left( \frac{2\pi}{\lambda} y \right)
\exp\left[ \lambda A {\rm e}^{-\frac{2\pi}{\lambda}x}
\cos\left( \frac{2\pi}{\lambda} y\right) \right] \nonumber \\ 
= \frac{\partial}{\partial x}
\exp\left[ \lambda A {\rm e}^{-\frac{2\pi}{\lambda}x}
\cos\left( \frac{2\pi}{\lambda} y\right) \right] ,  
\end{eqnarray}
the subsequent integration by parts together with the definition 
of the Bessel function (\ref{Bessel}) confirm the validity of
the contact value theorem (\ref{cvtsimple}).

In the case of the uniform line charge density $(A=0)$, taking into
account that $I_0(0)=1$, the formula (\ref{nxyfinal}) implies the particle
number density which depends only on the $x$-coordinate:
\begin{eqnarray}
n(x) & = & \frac{1}{4\pi} \int_0^{4\pi\sigma} {\rm d}s\, s {\rm e}^{-x s}
\nonumber \\ & = & \frac{1}{4\pi x^2}
\left[ 1 - \left( 1 + 4\pi\sigma x \right) {\rm e}^{-4\pi\sigma x} \right] ,
\qquad A=0 , \label{nx}
\end{eqnarray}
see also \cite{Jancovici84,Samaj13,Samaj20}.
The counterion density at the contact with the charged line
$n(0)=2\pi\sigma^2$ is in agreement with the contact value theorem
(\ref{cvtheorem}) taken at the coupling constant $\Gamma=2$.

The result (\ref{nxyfinal}) tells us that the particle number density at
the charged line ($x=0$) reads as 
\begin{eqnarray}
n(0,y) & = & \frac{1}{4\pi} \exp\left[ \lambda A
\cos\left( \frac{2\pi}{\lambda} y\right) \right] 
\int_0^{4\pi\sigma} {\rm d}s\,
\frac{1}{\int_0^{\infty} {\rm d}x'\,
{\rm e}^{-s x'} I_0\left( \lambda A {\rm e}^{-\frac{2\pi}{\lambda}x'} \right)} .
\nonumber \\ & & \label{n0y}
\end{eqnarray}
The mean value of the contact counterion density along the $y$-axis
is given by
\begin{equation}
\bar{n}(0) = \frac{1}{\lambda} \int_0^{\lambda} {\rm d}y\, n(0,y)
= \frac{1}{4\pi} \int_0^{4\pi\sigma} {\rm d}s\,
\frac{I_0(\lambda A)}{\int_0^{\infty} {\rm d}x'\,
{\rm e}^{-s x'} I_0\left( \lambda A {\rm e}^{-\frac{2\pi}{\lambda}x'} \right)} .  
\end{equation}
Provided that $\lambda\ne 0$, using the substitutions $x'=\lambda x''$
and $s=r/\lambda$ this relation can be rewritten in a dimensionless form
\begin{equation}
\lambda^2 \bar{n}(0) =
\frac{1}{4\pi} \int_0^{4\pi\lambda\sigma} {\rm d}r\,
\frac{I_0(\lambda A)}{\int_0^{\infty} {\rm d}x''\,
{\rm e}^{-r x''} I_0\left( \lambda A {\rm e}^{-2\pi x''} \right)} .  
\end{equation}
Here, the dimensionless quantity $\lambda^2 \bar{n}(0)$ is the function
of the dimensionless variables $\lambda\sigma$ and $\lambda A$;
the condition (\ref{cond}) restricts $\lambda\sigma\le 1$ and
the condition (\ref{inequality}) restricts
$\vert\lambda A\vert \le \lambda\sigma$.
Since the Bessel function $I_0(t)$ grows with increasing the argument $t$,
it holds that
$I_0(\lambda A) \ge I_0\left( \lambda A {\rm e}^{-2\pi x''} \right)$
for $0\le x''\le \infty$.
Consequently,
\begin{equation}
\bar{n}(0) \ge 2\pi \sigma^2
\end{equation}
where the equality takes place for the homogeneous line charge density
$(A=0)$ only.
We conclude that the variation of the line charge density around its mean value
$\sigma e$ induces an enhancement of the counterion density at the wall
in comparison with the line charged uniformly by $\sigma e$,
irrespective of the amplitude $A$ and the period $\lambda$ of
the cosine modulation.
This result, obtained in 2D at the finite coupling constant $\Gamma=2$,
is in agreement with the mean-field predictions and MC simulations
\cite{Lukatsky02a,Henle04,Fleck05}.

\begin{figure}[t]
\begin{center}
\includegraphics[width=0.7\textwidth,angle=-90,clip]{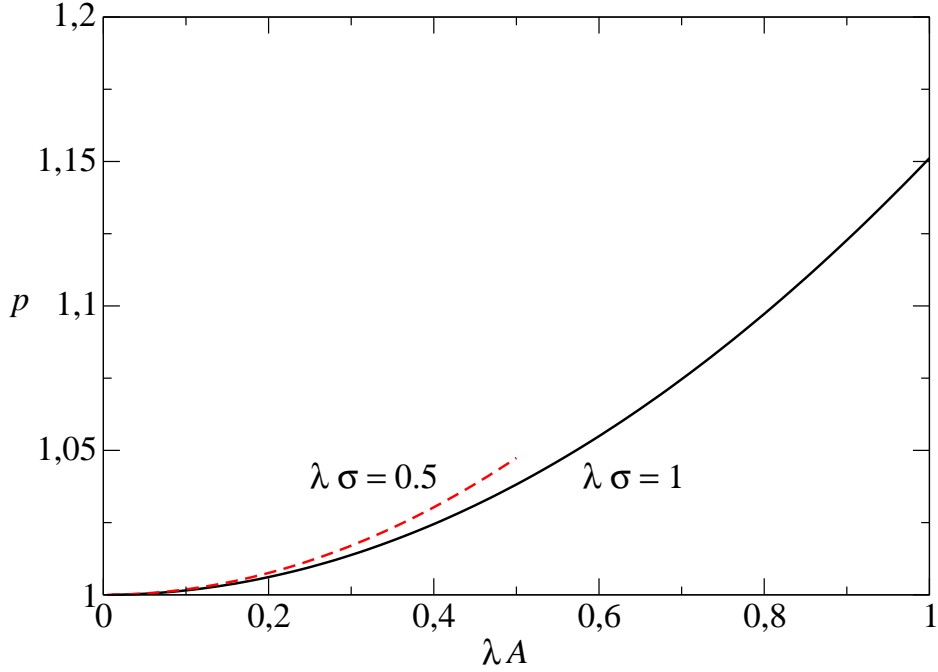}
\caption{The plots of the ratio (\ref{ratio}) ($p\ge 1$) versus
the dimensionless amplitude $\lambda A$ of the cosine modification of
the uniform line charge density $\sigma$.
The solid curve corresponds to $\lambda\sigma=1$ $(\lambda A \in [0,1])$
and the dashed curve to $\lambda\sigma=0.5$ $(\lambda A \in [0,0.5])$.}
\label{Fig:3}
\end{center}
\end{figure}

The numerical plots of the ratio
\begin{equation} \label{ratio}
p(\lambda\sigma,\lambda A) = \frac{\bar{n}(0)}{2\pi \sigma^2}
=  \frac{1}{8 (\pi\lambda\sigma)^2} \int_0^{4\pi\lambda\sigma} {\rm d}r\,
\frac{I_0(\lambda A)}{\int_0^{\infty} {\rm d}x''\,
{\rm e}^{-r x''} I_0\left( \lambda A {\rm e}^{-2\pi x''} \right)}  
\end{equation}  
versus the dimensionless amplitude $\lambda A$ of the
cosine modification (\ref{deltasigma}) of the uniform line charge density 
$\sigma$ are pictured in figure \ref{Fig:3} for two values of
the dimensionless $\lambda \sigma$.
The solid curve corresponds to $\lambda\sigma=1$ $(\lambda A \in [0,1])$
and the dashed curve to $\lambda\sigma=0.5$ $(\lambda A \in [0,0.5])$.
It is seen that the ratio $p$ is always greater than 1 as it should be. 
For a fixed value of $\lambda A$, the ratio $p$ increases with decreasing
$\lambda \sigma$.

\renewcommand{\theequation}{5.\arabic{equation}}
\setcounter{equation}{0}

\section{Effective interaction between two parallel EDLs} \label{Section5}

\subsection{Analogue of the contact value theorem} \label{Section5.1}
For the geometry of two parallel walls at distance $d$, one can proceed as
in the single wall case up to equation (\ref{crucialeq}).
Inserting there the equality
\begin{eqnarray}
\frac{\partial}{\partial x} \ln w(x,y) & = & - \frac{\pi\Gamma}{W}
+ 2\pi\Gamma\sigma_R + \Gamma \int_0^W {\rm d}y'\, \delta\sigma_L(y')
\frac{\partial \phi(x,y-y')}{\partial x} \nonumber \\ & & +  
\int_0^W {\rm d}y'\, \delta\sigma_R(y')
\frac{\partial \phi(d-x,y-y')}{\partial x} 
\end{eqnarray}  
holding for the one-body weight function $w(x,y)$ (\ref{onebody3}),
one gets
\begin{eqnarray}
\frac{W}{2\pi} \int_0^W {\rm d}y\,  \left[ n(0,y) - n(d,y) \right]
+ W\Gamma\sigma_R N + \frac{W \Gamma}{2\pi}
\int_0^W {\rm d}y \int_0^{\infty} {\rm d}x\, n(x,y)
\nonumber \\ \times \int_0^W {\rm d}y'\, \left[ \delta\sigma_L(y')
\frac{\partial \phi(x,y-y')}{\partial x} + \delta\sigma_R(y')
\frac{\partial \phi(d-x,y-y')}{\partial x} \right] = \frac{\Gamma}{2} N^2 .
\nonumber \\
\end{eqnarray}
Using the electroneutrality condition (\ref{LLRR}), this relation can be
expressed as
\begin{eqnarray}
\int_0^W {\rm d}y\,  \left[ n(0,y) - \pi \Gamma \sigma_L^2 \right]
\nonumber \\
+ \Gamma \int_0^W {\rm d}y \int_0^d {\rm d}x\, n(x,y)
\int_0^W {\rm d}y'\, \delta\sigma_L(y')
\frac{\partial \phi(x,y-y')}{\partial x} \nonumber \\
= \int_0^W {\rm d}y\,  \left[ n(d,y) - \pi \Gamma \sigma_R^2 \right]
\nonumber \\ - \Gamma \int_0^W {\rm d}y \int_0^d {\rm d}x\, n(x,y)
\int_0^W {\rm d}y'\, \delta\sigma_R(y')
\frac{\partial \phi(d-x,y-y')}{\partial x} . \label{equivalence}
\end{eqnarray}  
This equation defines specific combinations of integrals over the
particle and line charge densities which are invariant with respect
the left or right EDLs.
Note that in the limit $d\to\infty$, when the two EDLs decouple from
one another, both sides of equation (\ref{equivalence}) vanish
due to the contact value theorem (\ref{cvt}).

\subsection{The effective interaction} \label{Section5.2}
The pressure between the two lines is given by the
general formula (\ref{pressure}).
Considering the anticommuting-field representation of $Q_N$ (\ref{antipart}), 
the derivative of $Q_N$ (\ref{antipart}) with respect to the distance
$d$ can be expressed as follows
\begin{equation}
\frac{1}{Q_N} \frac{\partial Q_N}{\partial d} 
= \sum_{j,k=0}^{\gamma(N-1)} \langle \Xi_j \Psi_k \rangle
\frac{\partial w_{jk}}{\partial d} .
\end{equation}
For the interaction strengths (\ref{wjk}), defined in our case as
\begin{equation}
w_{jk} = \int_0^d {\rm d}x \int_0^W {\rm d}y\,
w(x,y) {\rm e}^{-\frac{2\pi}{W}j z} {\rm e}^{-\frac{2\pi}{W}k \bar{z}} ,  
\end{equation}  
considering the explicit form of the one-body weight function (\ref{onebody3})
it holds that
\begin{eqnarray}
\frac{\partial w_{jk}}{\partial d} & = &  \int_0^W {\rm d}y\,
w(d,y) {\rm e}^{-\frac{2\pi}{W}(j+k)d} {\rm e}^{-\frac{2\pi}{W}(j-k){\rm i}y}
\nonumber \\  & & + \int_0^d {\rm d}x \int_0^W {\rm d}y\,
w(x,y) {\rm e}^{-\frac{2\pi}{W}j z} {\rm e}^{-\frac{2\pi}{W}k \bar{z}}
\nonumber \\ & & \times
\Gamma \int_0^W {\rm d}y' \delta\sigma_R(y')
\frac{\partial\phi(d-x,y-y')}{\partial d} .
\end{eqnarray}  
Thus, the pressure is given by
\begin{eqnarray}
\beta P_N W & = & \int_0^W {\rm d}y\, \left[ n(d,y)-\pi\Gamma\sigma_R^2 \right]
\nonumber \\ & & -
\Gamma \int_0^d {\rm d}x \int_0^W {\rm d}y\, n(x,y)
\int_0^W {\rm d}y' \delta\sigma_R(y')
\frac{\partial\phi(d-x,y-y')}{\partial x} \nonumber \\ & & -
\Gamma \int_0^W {\rm d}y \int_0^W {\rm d}y'\,
\delta\sigma_L(y) \frac{\partial\phi(d,y-y')}{\partial d} 
\delta\sigma_R(y') .
\end{eqnarray}  
Taking into account the left-right equivalence relation (\ref{equivalence}),
this formula is equivalent to the one
\begin{eqnarray}
\beta P_N W & = & \int_0^W {\rm d}y\, \left[ n(0,y)-\pi\Gamma\sigma_L^2 \right]
\nonumber \\ & & +
\Gamma \int_0^d {\rm d}x \int_0^W {\rm d}y\, n(x,y)
\int_0^W {\rm d}y' \delta\sigma_L(y')
\frac{\partial\phi(x,y-y')}{\partial x} \nonumber \\ & & -
\Gamma \int_0^W {\rm d}y \int_0^W {\rm d}y'\,
\delta\sigma_L(y) \frac{\partial\phi(d,y-y')}{\partial d} 
\delta\sigma_R(y') .
\end{eqnarray}  

\subsection{The exactly solvable $\Gamma=2$ model: the particle number density}
\label{Section5.3}
Let the left and right (closed) lines be charged by the similarly modulated
line charge densities, up to the amplitude, say
\begin{equation} \label{linecharge}
\delta\sigma_L(y) = A_L \cos\left( \frac{2\pi}{\lambda} y \right) , \qquad
\delta\sigma_R(y) = A_R \cos\left( \frac{2\pi}{\lambda} y \right) ,  
\end{equation}
where $\lambda=W/M$ $(M=1,2,\ldots)$.
To ensure that the line charge densities $\sigma_L(y)\ge 0$ and
$\sigma_R(y)\ge 0$ for $y\in [0,W]$, the amplitudes are constrained by
\begin{equation}
\vert A_L\vert \le \sigma_L, \qquad \vert A_R\vert \le \sigma_R .
\end{equation}
We consider the free fermion coupling $\Gamma=2$.
Using that for $\delta\sigma(y)$ of the form (\ref{deltasigma})
the integral (\ref{Vxy}) results in (\ref{Vxyresult}), one has
\begin{equation}
2 \int_0^W {\rm d}y'\, \delta\sigma_L(y') \phi(x,y-y') =
A_L \lambda \exp\left( - \frac{2\pi}{\lambda} x \right)
\cos\left( \frac{2\pi}{\lambda} y \right)
\end{equation}
and
\begin{eqnarray}
2 \int_0^W {\rm d}y'\, \delta\sigma_R(y') \phi(d-x,y-y') & = &
A_R \lambda \exp\left[ - \frac{2\pi}{\lambda} (d-x) \right]
\cos\left( \frac{2\pi}{\lambda} y \right) . \nonumber \\ & &
\end{eqnarray}
The renormalized one-body weight function (\ref{onebody3}) then
takes the form 
\begin{eqnarray} 
w(x,y) & = & \frac{4\pi}{W^2} \exp\Bigg\{ - \frac{2\pi}{W} x
+ 4 \pi \sigma_R x \nonumber \\ & &
+ \lambda \left[ A_L {\rm e}^{-\frac{2\pi}{\lambda}x} +
A_R {\rm e}^{-\frac{2\pi}{\lambda}(d-x)} \right]
\cos\left( \frac{2\pi}{\lambda} y \right) \Bigg\} .
\end{eqnarray}

Since the function $w(x,y)$ is periodic along the $y$-axis with the period
$\lambda$, one can use the same arguments as in the semi-infinite
geometry to prove that under the condition
\begin{equation} \label{twocondition}
\lambda (\sigma_L + \sigma_R) \le 1  
\end{equation}
the matrix of interaction strengths (\ref{wjk}) is diagonal,
$w_{jk}=w_j\delta_{jk}$ with the positive diagonal elements
\begin{equation} \label{wdiag}
w_j = \frac{4\pi}{W} \int_0^d {\rm d}x\,
{\rm e}^{-\frac{4\pi}{W}(j+\frac{1}{2}-\sigma_R W)x}
I_0\left( \lambda\left[ A_L {\rm e}^{-\frac{2\pi}{\lambda}x} +
A_R {\rm e}^{-\frac{2\pi}{\lambda}(d-x)} \right] \right) .
\end{equation}
The particle number density (\ref{nxy}) reads as
\begin{eqnarray}
n(x,y) & = & \frac{4\pi}{W^2} \exp\left( \lambda
\left[ A_L {\rm e}^{-\frac{2\pi}{\lambda}x} +
A_R {\rm e}^{-\frac{2\pi}{\lambda}(d-x)} \right]   
\cos\left( \frac{2\pi}{\lambda} y\right) \right) \nonumber \\ 
& & \times \sum_{j=0}^{N-1} \frac{1}{w_j}
\exp^{-\frac{4\pi}{W} \left( j+\frac{1}{2}-\sigma_R W\right) x} .
\end{eqnarray}
In the thermodynamic limit $N,W\to\infty$ with the fixed ratio
$N/W = \sigma_L + \sigma_R$, introducing the continuous variable
$t=\frac{1}{N}\left( j+\frac{1}{2}-\sigma_R W \right)$,
transforming the sum $\frac{1}{N} \sum_{j=0}^{N-1}$ to the
integral over $t$ and using the substitution $s=4\pi(\sigma_L+\sigma_R)t$,
one gets
\begin{eqnarray}
n(x,y) & = & \frac{1}{4\pi} \exp\left( \lambda
\left[ A_L {\rm e}^{-\frac{2\pi}{\lambda}x} +
A_R {\rm e}^{-\frac{2\pi}{\lambda}(d-x)} \right]   
\cos\left( \frac{2\pi}{\lambda} y\right) \right) \nonumber \\ 
& & \times \int_{-4\pi\sigma_R}^{4\pi\sigma_L}
\frac{{\rm d}s\, {\rm e}^{-s x}}{\int_0^d {\rm d}x'\, {\rm e}^{- s x'}
I_0\left(\lambda \left[ A_L {\rm e}^{-\frac{2\pi}{\lambda}x'} +
A_R {\rm e}^{-\frac{2\pi}{\lambda}(d-x')} \right] \right)} . \label{nxyfinall}
\end{eqnarray}
Introducing the auxiliary function
\begin{eqnarray}
n_0(x,y;\sigma,\{ A,A'\}) & = & \frac{1}{4\pi} \exp\left( \lambda
\left[ A {\rm e}^{-\frac{2\pi}{\lambda}x} +
A' {\rm e}^{-\frac{2\pi}{\lambda}(d-x)} \right]   
\cos\left( \frac{2\pi}{\lambda} y\right) \right) \nonumber \\ 
& & \times \int_0^{4\pi\sigma}
\frac{{\rm d}s\, {\rm e}^{-s x}}{\int_0^d {\rm d}x'\, {\rm e}^{- s x'}
I_0\left(\lambda \left[ A {\rm e}^{-\frac{2\pi}{\lambda}x'} +
A' {\rm e}^{-\frac{2\pi}{\lambda}(d-x')} \right] \right)} \nonumber \\
& &
\end{eqnarray}
this formula can be rewritten as follows
\begin{equation} \label{particledensity}
n(x,y) = n_0(x,y;\sigma_L,\{ A_L,A_R\}) + n_0(d-x,y;\sigma_R,\{ A_R,A_L\}) .
\end{equation}
As before in the one-wall geometry, the particle number density is periodic
along the $y$-axis, with the same period $\lambda$ as the line charge
densities (\ref{linecharge}).
Over one period, it obeys the electroneutrality condition
\begin{equation}
\frac{1}{\lambda} \int_0^d {\rm d}x \int_0^{\lambda} {\rm d}y\,
n(x,y) = \sigma_L + \sigma_R .
\end{equation}

For the uniform line charge densities ($A_L=A_R=0$), 
one has $n_0(x,y;\sigma,\{ 0,0\}) \equiv n_0(x;\sigma)$ with
\begin{equation}
n_0(x;\sigma) = \frac{1}{4\pi} \int_0^{4\pi\sigma} {\rm d}s\, s
\frac{{\rm e}^{-sx}}{1-{\rm e}^{-sd}} .  
\end{equation}
According to (\ref{particledensity}), the particle number
density is given by
\begin{equation} 
n(x) = n_0(x;\sigma_L) + n_0(d-x;\sigma_R) ,
\end{equation}
in agreement with the previous result obtained in \cite{Samaj20}.

\subsection{The exactly solvable $\Gamma=2$ model: the pressure}
\label{Section5.4}
The pressure at $\Gamma=2$ will be calculated by using the formula
(\ref{pressure}),
\begin{eqnarray}
\beta P_N & = & -\frac{2}{W} \int_0^W {\rm d}y \int_0^W {\rm d}y'\,
\delta\sigma_L(y) \frac{\partial\phi(d,y-y')}{\partial d} \delta\sigma_R(y') 
\nonumber \\ & &
- 2 \pi\sigma_R^2 + \frac{\partial}{\partial d} \frac{1}{W} \ln Q_N .
\end{eqnarray}  
For the line charge deviations from the uniform densities (\ref{linecharge}),
it is simple to show that
\begin{equation}
-\frac{2}{W} \int_0^W {\rm d}y \int_0^W {\rm d}y'\,
\delta\sigma_L(y) \frac{\partial\phi(d,y-y')}{\partial d} \delta\sigma_R(y')
= \pi A_L A_R {\rm e}^{-\frac{2\pi}{\lambda}d} .
\end{equation}
For the diagonal matrix of interaction strengths with the nonzero elements
(\ref{wdiag}) it holds that $Q_N = \prod_{j=0}^{N-1} w_j$, i.e.,
\begin{eqnarray}
\frac{1}{W} \ln Q_N & = & \frac{N}{W} \frac{1}{N} \sum_{j=0}^{N-1}
\ln \Bigg\{ \frac{4\pi}{W} \int_0^d {\rm d}x\,
{\rm e}^{-\frac{4\pi}{W}(j+\frac{1}{2}-\sigma_R W)x} \nonumber \\ & & \times
I_0\left( \lambda\left[ A_L {\rm e}^{-\frac{2\pi}{\lambda}x} +
A_R {\rm e}^{-\frac{2\pi}{\lambda}(d-x)} \right] \right) \Bigg\} .
\end{eqnarray}  
Going to the thermodynamic limit $N,W\to\infty$ with the fixed
ratio $N/W=\sigma_L+\sigma_R$ by converting the discrete sum to
the continuous integral and forgetting the irrelevant term
$\propto \ln(4\pi/W)$ which does not depend on the distance $d$, one gets
\begin{eqnarray}
\lim_{N,W\to\infty} \frac{1}{W} \ln Q_N & = & \frac{1}{4\pi}
\int_{-4\pi\sigma_R}^{4\pi\sigma_L} {\rm d}s\, \ln \Bigg\{
\int_0^d {\rm d}x\, {\rm e}^{-sx} \nonumber \\ & & \times
I_0\left( \lambda\left[ A_L {\rm e}^{-\frac{2\pi}{\lambda}x} +
A_R {\rm e}^{-\frac{2\pi}{\lambda}(d-x)} \right] \right) \Bigg\} .
\label{while}
\end{eqnarray}  
Introducing the auxiliary function
\begin{eqnarray}
f(d,\sigma;\{A,A'\}) & = & \frac{1}{4\pi} \int_0^{4\pi\sigma} {\rm d}s\,
\ln \Bigg\{ \int_0^d {\rm d}x\, {\rm e}^{-sx} \nonumber \\ & & \times
I_0\left( \lambda\left[ A {\rm e}^{-\frac{2\pi}{\lambda}x} +
A' {\rm e}^{-\frac{2\pi}{\lambda}(d-x)} \right] \right) \Bigg\} , \label{ff}   
\end{eqnarray}  
the expression (\ref{while}) can be written in a symmetric form
\begin{equation}
\lim_{N,W\to\infty} \frac{1}{W} \ln Q_N = 2\pi\sigma_R^2 d
+ f(d,\sigma_L;\{A_L,A_R\}) + f(d,\sigma_R;\{A_R,A_L\}) .
\end{equation}
Consequently,
\begin{eqnarray} 
\beta P & = & \pi A_L A_R {\rm e}^{-2\pi d/\lambda}
+ \frac{\partial}{\partial d} f(d,\sigma_L;\{A_L,A_R\}) \nonumber \\
& & + \frac{\partial}{\partial d} f(d,\sigma_R;\{A_R,A_L\}) .  
\label{Pressure}
\end{eqnarray}  
Provided that $\lambda\ne 0$, using the substitutions $x=\lambda x'$
and $s=r/\lambda$ in the definition (\ref{ff}) of $f(d,\sigma;\{A,A'\})$
and introducing the dimensionless pressure
\begin{equation}
\widetilde{P}\equiv \lambda^2 \beta P ,
\end{equation}
this relation can be rewritten
in a dimensionless form
\begin{eqnarray} 
\widetilde{P} & = & \pi (\lambda A_L) (\lambda A_R)
{\rm e}^{-2\pi d/\lambda} + \frac{\partial}{\partial (d/\lambda)} 
g\left( d/\lambda,\lambda\sigma_L;\{\lambda A_L,\lambda A_R\} \right)
\nonumber \\ & & + \frac{\partial}{\partial (d/\lambda)} 
g\left( d/\lambda,\lambda\sigma_R;\{\lambda A_R,\lambda A_L\} \right),
\label{Pressureprime}
\end{eqnarray}
where
\begin{eqnarray}
g\left( d/\lambda,\lambda\sigma;\{\lambda A,\lambda A'\} \right)
& = & \frac{1}{4\pi} \int_0^{4\pi\lambda\sigma} {\rm d}r\,
\ln \Bigg\{ \int_0^{d/\lambda} {\rm d}x'\, {\rm e}^{-r x'} \nonumber \\ & & \times
I_0\left( \lambda A {\rm e}^{-2\pi x'} + \lambda A'
{\rm e}^{-2\pi d/\lambda+2\pi x'} \right) \Bigg\} . \label{fff}   
\end{eqnarray}  

In the case of the uniform line charge densities $(A_L=A_R=0)$,  
using that $I_0(0)=1$ this formula results in the pressure
\begin{equation} \label{press}
\beta P_0(d;\{ \sigma_L,\sigma_R\})
= \beta {\cal P}(d,\sigma_L) + \beta {\cal P}(d,\sigma_R) ,
\end{equation}
where
\begin{equation} 
\beta {\cal P}(d,\sigma)
= \frac{1}{2\pi d^2} \int_0^{2\pi\sigma d} {\rm d}t\,
\frac{t}{\sinh t} {\rm e}^{-t}
= \frac{1}{2\pi} \int_0^{2\pi\sigma} {\rm d}s\,
\frac{s}{\sinh(d s)} {\rm e}^{-d s}  
\end{equation}  
is the (positive) pressure between two parallel lines at distance $d$,
the one with the uniform line charge density $\sigma$ and
the other with zero line charge density; this result coincides with that
in \cite{Samaj20}.
Since
\begin{equation}
\frac{\partial}{\partial d} \beta {\cal P}(d,\sigma)
= - \frac{1}{2\pi} \int_0^{2\pi\sigma} {\rm d}s\,
\left[ \frac{s}{\sinh(d s)} \right]^2 < 0 , 
\end{equation}
the positive pressure $P_0$ decays monotonously from $\infty$ at $d\to 0$
to 0 at $d\to\infty$, i.e., two likely charged lines
with $\sigma_L,\sigma_R>0$ always repel one another.

For the lines with inhomogeneous line charge densities, we restrict ourselves
to the symmetric mean line charge densities
\begin{equation} \label{LR}
\sigma_L = \sigma_R \equiv \sigma . 
\end{equation}
The condition (\ref{twocondition}) is equivalent to the one
$\lambda \sigma \le \frac{1}{2}$; to simplify the analysis the value
of $\lambda \sigma$ is set to its maximum:
\begin{equation}
\lambda\sigma = \frac{1}{2} .
\end{equation}
As concerns the amplitudes of the cosine modulations of the line charge
densities (\ref{deltasigma}), two choices are considered.

\begin{figure}[t]
\begin{center}
\includegraphics[width=0.7\textwidth,angle=-90,clip]{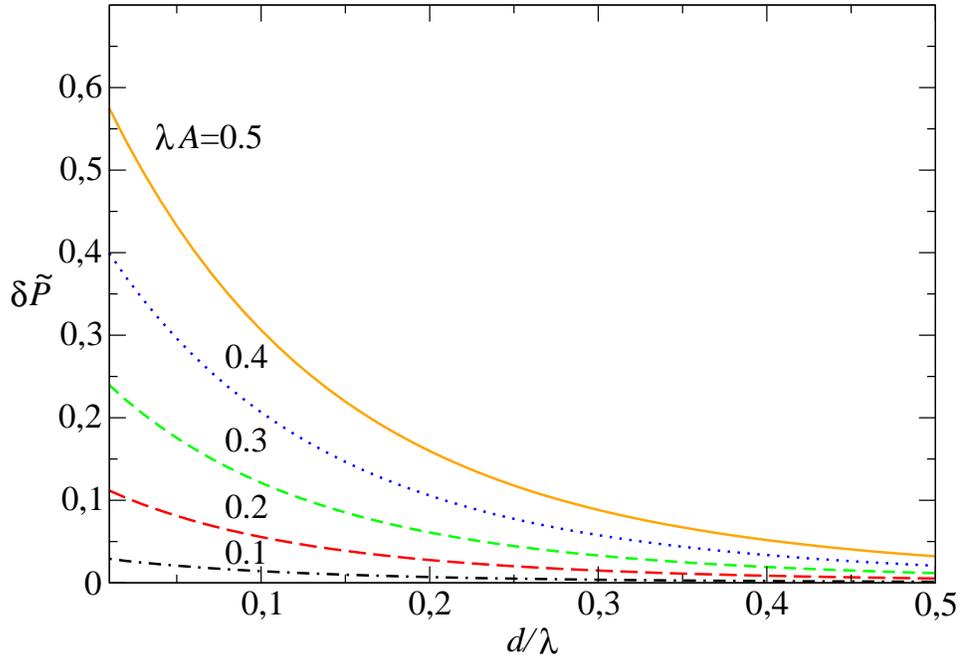}
\caption{The plots of the (positive) difference between the rescaled
pressures for uniformly and periodically (period $\lambda$) charged lines
(\ref{pressdif}) with the same mean line charge density given by
$\lambda\sigma=\frac{1}{2}$
versus the dimensionless distance between the lines $d/\lambda$.  
The equivalent left and right charge densities are in phase with respect
to one another.
The curves correspond from up to down to five values of the
(dimensionless) modulation amplitude $\lambda A = 0.5, 0.4, 0.3, 0.2, 0.1$.} 
\label{Fig:4}
\end{center}
\end{figure}

The choice of the equivalent (positive) amplitudes
\begin{equation}
A_L = A_R \equiv A , \qquad \lambda A \le \lambda\sigma = \frac{1}{2} ,
\end{equation}  
describes the identical left and right line charge densities which are
in phase with respect to one another.
The rescaled pressure (\ref{Pressureprime}) is now given by
\begin{equation} 
\widetilde{P}\left( d/\lambda,\lambda A \right)
= \pi (\lambda A)^2 {\rm e}^{-2\pi d/\lambda}
+ 2 \frac{\partial}{\partial (d/\lambda)} 
g\left( d/\lambda,\lambda A \right) ,
\end{equation}
where
\begin{eqnarray}
g\left( d/\lambda,\lambda A \right) & = & \frac{1}{4\pi} \int_0^{2\pi} {\rm d}r\,
\ln \Bigg\{ \int_0^{d/\lambda} {\rm d}x'\, {\rm e}^{-r x'} \nonumber \\ & & \times
I_0\left[ \lambda A \left( {\rm e}^{-2\pi x'} + {\rm e}^{-2\pi d/\lambda+2\pi x'}
\right) \right] \Bigg\} .   
\end{eqnarray}  
The numerical treatment of this result implies that, as in the case of
uniformly charged lines, the pressure $\widetilde{P}$ is always positive
and decays monotonously from $\infty$ at $d/\lambda\to 0$ to 0
at $d/\lambda\to\infty$.
It turns out that at arbitrary distance $d/\lambda$ the pressure
$\widetilde{P}$ is always smaller than the rescaled pressure of
the uniformly charged lines $\sigma_L=\sigma_R=\sigma$ with
$\lambda\sigma=\frac{1}{2}$,
\begin{equation} \label{P0}
\lambda^2 \beta P_0(d/\lambda)
= \frac{1}{\pi} \left( \frac{\lambda}{d} \right)^2
\int_0^{\pi d/\lambda} {\rm d}t\, \frac{t}{\sinh t} {\rm e}^{-t} .
\end{equation}
The (positive) difference between the rescaled pressures 
\begin{equation} \label{pressdif}
\delta\widetilde{P}\left( d/\lambda,\lambda A \right) \equiv
\lambda^2 \beta P_0(d/\lambda)
- \widetilde{P}\left( d/\lambda,\lambda A \right) 
\end{equation}
is pictured in figure \ref{Fig:4} for five values of the modulation parameter
(from up to down) $\lambda A = 0.5, 0.4, 0.3, 0.2, 0.1$.
It is seen that the pressure difference increases with increasing the
dimensionless cosine amplitude of the modulated line charge density 
$\lambda A$ and with decreasing the dimensionless distance between the lines
$d/\lambda$.

\begin{figure}[t]
\begin{center}
\includegraphics[width=0.7\textwidth,angle=-90,clip]{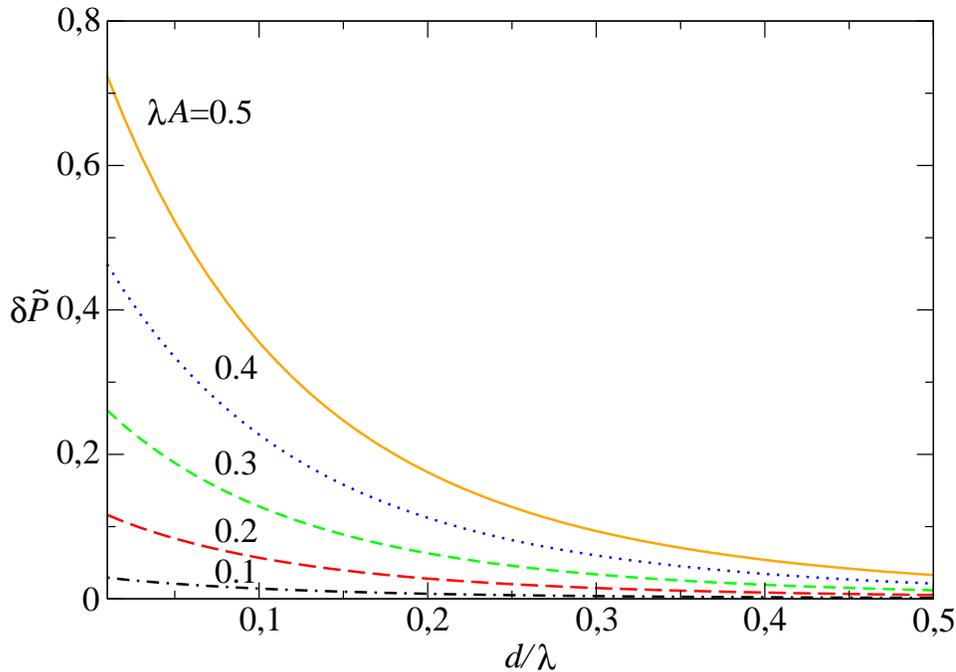}
\caption{The same as in figure \ref{Fig:4}, but the equivalent left and right
charge densities are shifted by a half-period $\lambda/2$ with respect
to one another.} 
\label{Fig:5}
\end{center}
\end{figure}

The choice of the amplitudes with opposite signs
\begin{equation}
A_L = - A_R \equiv A , \qquad \lambda A \le \frac{1}{2} ,
\end{equation}  
describes the identical left and right line charge densities which are
shifted in phase with respect to one another by a half-period $\lambda/2$.
The rescaled pressure (\ref{Pressureprime}) is given by
\begin{equation} 
\widetilde{P}\left( d/\lambda,\lambda A \right)
= - \pi (\lambda A)^2 {\rm e}^{-2\pi d/\lambda}
+ 2 \frac{\partial}{\partial (d/\lambda)} 
g\left( d/\lambda,\lambda A \right) ,
\end{equation}
where
\begin{eqnarray}
g\left( d/\lambda,\lambda A \right) & = & \frac{1}{4\pi} \int_0^{2\pi} {\rm d}r\,
\ln \Bigg\{ \int_0^{d/\lambda} {\rm d}x'\, {\rm e}^{-r x'} \nonumber \\ & & \times
I_0\left[ \lambda A \left( {\rm e}^{-2\pi x'} - {\rm e}^{-2\pi d/\lambda+2\pi x'}
\right) \right] \Bigg\} .   
\end{eqnarray}  
As before, the pressure $\widetilde{P}$ is positive
and decays monotonously from $\infty$ at $d/\lambda\to 0$ to 0
at $d/\lambda\to\infty$.
It is always smaller than the rescaled pressure of
the uniformly charged lines (\ref{P0}) taken at $\lambda\sigma=\frac{1}{2}$.
The positive difference between the pressures (\ref{pressdif})
is pictured in figure \ref{Fig:5} for five values of the parameter
$\lambda A = 0.5, 0.4, 0.3, 0.2, 0.1$.
In comparison with the results in figure \ref{Fig:4}, the half-period shift in
the line charge densities induces, for fixed values of the parameters
$\lambda A$ and $d/\lambda$, larger differences between the pressures.

We conclude that, at any distance between two lines, the modulation of
the line charge densities always leads to the diminution of the pressure
in comparison with the symmetric lines charged uniformly by the same mean
line charge densities.

\renewcommand{\theequation}{6.\arabic{equation}}
\setcounter{equation}{0}

\section{Conclusion} \label{Section6}
In this paper, the system of mobile charges is put on
the surface of the 2D cylinder (section \ref{Section2}).
The particles are confined to either the semi-infinite space bounded by
one circle or the finite space between two circles at distance $d$,
in both cases the circles are charged inhomogeneously by a fixed
line charge density.
The question is how the non-uniformity of the line charge densities
influences the particle density and the effective interaction between
two circles.

2D one-component Coulomb fluids in thermal equilibrium have the advantage
of being exactly solvable not only in the high-temperature PB limit of
the coupling constant $\Gamma\to 0$ but also, in principle,
at the finite free-fermion coupling $\Gamma=2$.
Moreover, for the series of coupling constants $\Gamma=2\gamma$
with $\gamma$ being a positive integer, the partition function and
the many-body particle densities of the 2D Coulomb system are expressible
explicitly in terms of the 1D anticommuting field theory defined on a chain
of sites (section \ref{Section3}).
Specific transformations of anticommuting variables, which keep
the composite form of the operators (\ref{composite}), lead to specific
sum rules, i.e., exact constraints for correlators of the composite operators.

As concerns the semi-infinite geometry of one EDL (section \ref{Section4}), 
the derived sum rule for correlators of the composite operators implies
the contact value theorem (\ref{cvt}) valid for any $\Gamma$.
Its form, which involves the particle density profile inside
the whole semi-infinite region accessible to particle, is much more
complicated that the one of the homogeneously charged line (\ref{cvtheorem}),
which involves only the particle density at the line contact.
At the free fermion coupling constant $\Gamma=2$, choosing the cosine
modulation of the line charge density with period $\lambda$ (\ref{deltasigma}),
the matrix of interaction strengths between the anticommuting variables becomes
diagonal under the condition (\ref{cond}) for the mean line charge density
$\sigma$.
This leads to the explicit form of the particle density
profile (\ref{nxyfinal}).
The numerical results for the ratio of the mean (averaged over the $y$-axis)
values of the contact densities of counterions for the modulated and
uniform line charge densities, always larger than 1 as is seen in
figure \ref{Fig:3}, confirm the anticipated enhancement of the counterion
density at the line due to the line charge modulation. 

The effective interaction (the pressure) between two EDLs at distance $d$
is discussed in section \ref{Section5}.
For the cosine modulation of left and right line charge densities
(\ref{linecharge}), under the condition (\ref{twocondition}) for
the left and right mean values of the line charge densities,
the pressure at the free fermion coupling $\Gamma=2$ is given
by equations (\ref{ff}) and (\ref{Pressure}).
Restricting ourselves to the symmetric left and right mean line
charge densities (\ref{LR}) with $\lambda\sigma=\frac{1}{2}$,
two choices for the amplitudes of the cosine modulations are considered:
$A_L=A_R\equiv A$ (the modulations are in phase, for numerical
results see figure \ref{Fig:4}) and $A_L=-A_R\equiv A$
(the modulations are shifted by half-period $\lambda/2$, for numerical
results see figure \ref{Fig:5}).
In both cases, similarly as in the uniform case, the positive pressure
is a decreasing function of the dimensionless distance between lines
$d/\lambda$.
In other words, the considered modulations of the line charge densities
are not strong enough to induce an effective attraction between the circles
in an interval of the distance $d/\lambda$. 
This feature does not contradict a recent study \cite{Samaj20}
dealing with finite-$N$ counterion systems on the surface of the cylinder
with symmetric uniformly charged circles at couplings $\Gamma=2,3,4$ which
indicates that the attraction arises somewhere between
$\Gamma=2$ and $\Gamma=4$.
As is seen in figures \ref{Fig:4} and \ref{Fig:5}, the modulation
of the line charge densities always decreases the pressure, which is
in agreement with the expectation from the mean-field PB theory
and MC simulations.

\ack
The support received from VEGA Grant No. 2/0092/21 and Project
APVV-20-0150 is acknowledged.


\section*{References}

\begin{thebibliography}{10}

\bibitem{Martin88} Martin Ph A 1988
Sum rules in charged fluids
{\it Rev. Mod. Phys.} {\bf 60} 1075--1127

\bibitem{Andelman06} Andelman D 2006
Introduction to electrostatics in soft and biological matter
In: Poon, W.C.K., Andelman, D. (eds.) {\it Soft Condensed Matter Physics
  in Molecular and Cell Biology} vol. 6.
(Taylor \& Francis, New York)

\bibitem{Levin02} Levin Y 2002
Electrostatic correlations: from Plasma to Biology
{\it Rep. Prog. Phys.} {\bf 65} 1577--1632 

\bibitem{Gulbrand84} Gulbrand L, J\"onsson B, Wennerstr\"om H and Linse P 1984 
Electrical double layer forces. A Monte Carlo study
{\it J. Chem. Phys.} {\bf 80} 2221-2228

\bibitem{Attard88} Attard P, Mitchell D J and Ninham B W 1988
Beyond Poisson-Boltzmann: Images and correlations in the electric double
layer. I. Counterions only
{\it J. Chem. Phys.} {\bf 88} 4987--4996

\bibitem{Attard96} Attard Ph 1996
Electrolytes and the electric double layer
{\it Adv. Chem. Phys.} {\bf XCII} 1--159

\bibitem{Messina09} Messina R 2009
Electrostatics in soft matter
{\it J. Phys.: Condens. Matter} {\bf 21} 113102

\bibitem{Henderson78} Henderson D and Blum L 1978 
Some exact results and the application of the mean spherical approximation 
to charged hard spheres near a charged hard wall
{\it J. Chem. Phys.} {\bf 69} 5441--5449

\bibitem{Henderson79} Henderson D, Blum L and Lebowitz J L 1979
An exact formula for the contact value of the density profile of 
a system of charged hard spheres near a charged wall
{\it J. Electroanal. Chem.} {\bf 102} 315--319

\bibitem{Blum81} Blum L, Henderson D, Lebowitz J L, Gruber Ch and 
Martin Ph A 1981 
A sum rule for an inhomogeneous electrolyte
{\it J. Chem. Phys.} {\bf 75} 5974--5975 

\bibitem{Carnie81} Carnie S L and Chan D Y C 1981 
The Stillinger-Lovett condition for non-uniform electrolytes
{\it Chem. Phys. Lett.} {\bf 77} 437--440

\bibitem{Hansen00} Hansen J P and L\"owen H 2000
Effective interactions between electric double layers
{\it Annu. Rev. Phys. Chem.} {\bf 51} 209--242 

\bibitem{Khan85} Khan A, J\"onsson B and Wennerstr\"om H 1985 
Phase equilibria in the mixed sodium and calcium di-2-ethylhexylsulfosuccinate 
aqueous system. An illustration of repulsive and attractive double-layer forces
{\it J. Phys. Chem.} {\bf 89} 5180--5184

\bibitem{Kjellander88} Kjellander R, Mar\v{c}elja S and Quirk J P 1988 
Attractive double-layer interactions between calcium clay particles
{\it J. Colloid Interface Sci.} {\bf 126} 194--211

\bibitem{Bloomfield91} Bloomfield V A 1991 
Condensation of DNA by multivalent cations: Considerations on mechanism
{\it Biopolymers} {\bf 31} 1471--1481

\bibitem{Kekicheff93} K\'ekicheff P, Mar\v{c}elja S, Senden T J and 
Shubin V E 1993 
Charge reversal seen in electrical double layer interaction of surfaces 
immersed in 2:1 calcium electrolyte 
{\it J. Chem. Phys.} {\bf 99} 6098--6113

\bibitem{Dubois98} Dubois M, Zemb T, Fuller N, Rand R P and 
Pargesian V A 1998
Equation of state of a charged bilayer system: Measure of the entropy of 
the lamellar–lamellar transition in DDABr 
{\it J. Chem. Phys.} {\bf 108} 7855--7869

\bibitem{Kjellander84} Kjellander R and Mar\v{c}elja S 1984 
Correlation and image charge effects in electric double-layers
{\it Chem. Phys. Lett.} {\bf 112} 49--53

\bibitem{Gronbech97} Gr{\o}nbech-Jensen N, Mashl R J, Bruinsma R F and
Gelbart W M 1997 
Counterion-Induced attraction between rigid polyelectrolytes
{\it Phys. Rev. Lett.} {\bf 78} 2477--2480 

\bibitem{Boroudjerdi05} Boroudjerdi H, Kim Y-W, Naji A, Netz R R, 
Schlagberger X and Serr A 2005
Statics and dynamics of strongly charged soft matter
{\it Phys. Rep.} {\bf 416} 129--199

\bibitem{Naji13} Naji A, Kandu\v{c} M, Forsman J and Podgornik R 2013
Perspective: Coulomb fluids -- Weak coupling, strong coupling, in between and 
beyond
{\it J. Chem. Phys.} {\bf 139} 150901

\bibitem{Netz00} Netz R R, Orland H 2000
Beyond Poisson-Boltzmann: Fluctuation effects and correlation functions
{\it Eur. Phys. J.} E {\bf 1} 203--214

\bibitem{Podgornik90} Podgornik R 1990
An analytic treatment of the first-order correction to the Poisson-Boltzmann
interaction free energy in the case of counter-ion only Coulomb fluid
{\it J. Phys. A: Math. Gen.} {\bf 23} 275--284

\bibitem{Moreira00} Moreira A G and Netz R R 2000 
Strong-coupling theory for counter-ion distributions
{\it Europhys. Lett.} {\bf 52} 705--711

\bibitem{Moreira01} Moreira A G and Netz R R 2001 
Binding of similarly charged plates with counterions only
{\it Phys. Rev. Lett.} {\bf 87} 078301

\bibitem{Netz01} Netz R R 2001
Electrostatics of counter-ions at and between planar charged walls: from
Poisson-Boltzmann to the strong-coupling theory
{\it Eur. Phys. J.} E {\bf 5} 557--574

\bibitem{Moreira02} Moreira A G and Netz R R 2002 
Simulations of counterions at charged plates
{\it Eur. Phys. J.} E {\bf 8} 33--58

\bibitem{Kanduc07} Kandu\v{c} M and Podgornik R 2007 
Electrostatic image effects for counterions between charged planar walls
{\it Eur. Phys. J.} E {\bf 23} 265--274

\bibitem{Shklovskii99} Shklovskii B I 1999 
Screening of a macroion by multivalent ions: 
Correlation-induced inversion of charge
{\it Phys. Rev.} E {\bf 60} 5802--5811

\bibitem{Levin99} Levin Y, Arenzon J J and Stilck J F 1999 
The nature of attraction between like-charged rods
{\it Phys. Rev. Lett.} {\bf 83} 2680--2680

\bibitem{Grosberg02} Grosberg A Y, Nguyen T T and Shklovskii B I 2002 
Colloquium: The physics of charge inversion in chemical and biological systems
{\it Rev. Mod. Phys.} {\bf 74} 329--345

\bibitem{Samaj11a} \v{S}amaj L and Trizac E 2011
Counterions at highly charged interfaces: From one plate to like-charge
attraction
{\it Phys. Rev. Lett.} {\bf 106} 078301

\bibitem{Samaj11b} \v{S}amaj L and Trizac E 2011
Wigner-crystal formulation of strong-coupling theory for counterions
near planar charged interfaces
{\it Phys. Rev.} E {\bf 24} 041401

\bibitem{Nordholm84} Nordholm S 1984
Simple analysis of the thermodynamic properties of the one-component plasma
{\it Chem. Phys. Lett.} {\bf 105} 302--307

\bibitem{Forsman04} Forsman J 2004 
A simple correlation-corrected Poisson-Boltzmann theory
{\it J. Phys. Chem.} B {\bf 108} 9236--9245

\bibitem{Samaj16} \v{S}amaj L, dos Santos A P, Levin Y and Trizac E 2016 
Mean-field beyond mean-field: the single particle view for moderately 
to strongly coupled charged fluids
{\it Soft Matter} {\bf 12} 8768--8773

\bibitem{Palaia18} Palaia I, Trulsson M, \v{S}amaj L and Trizac E 2018
A correlation-hole approach to the electric double layer with counter-ions
only
{\it Mol. Phys.} {\bf 116} 3134--3146

\bibitem{Israelachvili92} Israelachvili J N 1992
{\it Intermolecular and Surface Forces} 
3rd edn. (Academic Press, San Diego)  

\bibitem{Leckband93} Leckband D E, Helm C A and Israelachvili J 1993
The role of calcium in the adhesion and fusion of mixed lipid bilayers
{\it Biochem.} {\bf 32} 1127--1140

\bibitem{Walz98} Walz J Y 1998
The effect of surface heterogeneities on colloidal forces
{\it Adv. Colloid Interface Sci.} {\bf 74} 119--168

\bibitem{Chan80} Chan D Y C, Mitchell J and Ninham B W 1980
A self‐consistent study of ion adsorption and discrete charge effects in
the electrical double layer
{\it J. Chem. Phys.} {\bf 72} 5159--5162

\bibitem{Kjellander88b} Kjellander R and Mar\v{c}elja S 1988
Inhomogeneous Coulomb fluids with image interactions between planar surfaces.
III. Distribution functions
{\it J. Chem. Phys.}  {\bf 88} 7138--7146

\bibitem{Lukatsky02a} Lukatsky D B, Safran S A, Lau A W C and
Pincus P A 2002
Enhanced counterion localization induced by surface charge modulation
{\it Europhys. Lett.} {\bf 58} 785--791

\bibitem{Henle04} Henle M L, Santangelo C D, Patel D M and Pincus P A 2004
Distribution of counterions near discretely charged planes and rods
{\it Europhys. Lett.} {\bf 66} 284--290

\bibitem{Fleck05} Fleck C C and Netz R. R. 2005
Counterions at disordered charged planar surfaces 
{\it Europhys. Lett.} {\bf 70} 341--347
  
\bibitem{Lukatsky02b} Lukatsky D B and Safran S A 2002 
Universal reduction of pressure between charged surfaces by
long-wavelength surface charge modulation
{\it Europhys. Lett.} {\bf 60} 629--635

\bibitem{Khan05} Khan M O, Petris S and Chan D Y C 2005
The influence of discrete surface charges on the force between charged surfaces
{\it J. Chem. Phys.} {\bf 122} 104705

\bibitem{Samaj19} \v{S}amaj L and Trizac E 2019
Electric double layers with surface charge modulations: Exact
Poisson-Boltzmann solutions
{\it Phys. Rev.} E {\bf 100} 042611

\bibitem{Jancovici81} Jancovici B 1981
Exact results for the two-dimensional one-component plasma
{\it Phys. Rev. Lett.} {\bf 46} 386--388

\bibitem{Alastuey81} Alastuey A and Jancovici B 1981 
On the classical two-dimensional one-component Coulomb plasma
{\it J. Physique} {\bf 42} 1--12

\bibitem{Jancovici92} Jancovici B 1992
Inhomogeneous two-dimensional plasmas
In: {\it Inhomogeneous Fluids} pp. 201-237 Henderson D (ed.) 
(Dekker, New York)

\bibitem{Forrester98} Forrester P J 1998 
Exact results for two-dimensional Coulomb systems
{\it Phys. Rep.} {\bf 301} 235--270  

\bibitem{Tellez99} T\'ellez G and Forrester P J 1999
Exact finite-size study of the 2d-OCP at $\Gamma=4$ and $\Gamma=6$ 
{\it J. Stat. Phys.} {\bf 97} 489--521

\bibitem{Tellez12} T\'ellez G and Forrester P J 2012 
Expanded Vandermonde powers and sum rules for the two-dimensional 
one-component plasma
{\it J. Stat. Phys.} {\bf 147} 825--855

\bibitem{Samaj95} \v{S}amaj L and Percus J K 1995
A functional relation among the pair correlations of the two-dimensional
one-component plasma
{\it J. Stat. Phys.} {\bf 80} 811--824

\bibitem{Samaj04a} \v{S}amaj L 2004
Is the two-dimensional one-component plasma exactly solvable?
{\it J. Stat. Phys.} {\bf 117} 131--158

\bibitem{Samaj04b} \v{S}amaj L, Wagner J and Kalinay P 2004 
Translation symmetry breaking in the one-component plasma on the cylinder
{\it J. Stat. Phys.} {\bf 117} 159--178

\bibitem{Samaj15} \v{S}amaj L 2015
Counter-ions near a charged wall: Exact results for disc and planar geometries
{\it J. Stat. Phys.} {\bf 161} 227--249

\bibitem{Jancovici84} Jancovici B 1984
Surface properties of a classical two-dimensional one-component plasma: 
Exact results
{\it J. Stat. Phys.} {\bf 34} 803--815

\bibitem{Samaj13} \v{S}amaj L 2013
Counter-ions at single charged wall: Sum rules
{\it Eur. Phys. J. E} {\bf 36} 100 

\bibitem{Samaj20} \v{S}amaj L 2020
Attraction of like-charged walls with counterions only: Exact results
for the 2D cylinder geometry  
{\it J. Stat. Phys.} {\bf 181} 1699--1729

\bibitem{Choquard81} Choquard Ph 1981 
The two-dimensional one component plasma on a periodic strip
{\it Helv. Phys. Acta} {\bf 54} 332--332  

\bibitem{Samaj14} \v{S}amaj L and Trizac E 2014
Counter-ions between or at asymmetrically charged walls: 2D free-fermion point
{\it J. Stat. Phys.} {\bf 156} 932--947

\bibitem{Samaj00} \v{S}amaj L 2000
Microscopic calculation of the dielectric susceptibility tensor for 
Coulomb fluids
{\it J. Stat. Phys.} {\bf 100} 949--967

\bibitem{Gradshteyn} Gradshteyn I S and Ryzhik I M 2000 
{\it Table of Integrals, Series, and Products} 
6th edn. (Academic Press, London)

\end{thebibliography}
\end{document}